\documentclass[12pt,letterpaper]{JHEP3}

\usepackage{amscd,amsmath,amssymb,amsfonts,xspace,mathrsfs}
\usepackage{epsfig}
\usepackage{bbm}
\usepackage{braket}
\usepackage{float}
\usepackage{euscript}
\usepackage{mathtools}
\usepackage{slashed}
\allowdisplaybreaks

\numberwithin{equation}{section}

\def\bea{\begin{eqnarray}}
\def\eea{\end{eqnarray}}
\def\be{\begin{equation}}
\def\ee{\end{equation}}
\def\ba{\begin{align}}
\def\ea{\end{align}}
\def\bse{\begin{subequations}}
\def\ese{\end{subequations}}

\newcommand{\Tr}{\mbox{Tr}}

\def\Im{\,{\rm Im}\,}
\def\Re{\,{\rm Re}\,}

\def\({\left(}
\def\){\right)}
\def\[{\left[}
\def\]{\right]}
\def\<{\left\langle}
\def\>{\right\rangle}
\def\hf{{1\over 2}}

\newcommand{\CP}{\IC P^1}

\newcommand{\p}{\partial}

\newcommand{\nn}{\nonumber}

\newcommand{\eps}{\epsilon}

\def\vph{\varphi}

\newcommand{\de}{\mathrm{d}}

\newcommand{\I}{\mathrm{i}}

\newcommand{\cL}{\mathcal{L}}
\newcommand{\cD}{\mathcal{D}}

\newcommand{\cC}{\mathcal{C}}
\newcommand{\cS}{\mathcal{S}}

\newcommand{\cK}{\mathcal{K}}
\newcommand{\cM}{\mathcal{M}}

\newcommand{\cN}{\mathcal{N}}

\newcommand{\cX}{\mathcal{X}}

\newcommand{\cR}{\mathcal{R}}

\newcommand{\cJ}{\mathcal{J}}
\newcommand{\cY}{\mathcal{Y}}
\newcommand{\cZ}{\mathcal{Z}}
\newcommand{\cI}{\mathcal{I}}

\newcommand{\cA}{\mathcal{A}}
\newcommand{\cB}{\mathcal{B}}

\newcommand{\scJ}{\mathscr{J}}
\newcommand{\scK}{\mathscr{K}}
\newcommand{\scL}{\mathscr{L}}

\newcommand{\bS}{\bar S}

\newcommand{\IR}{\mathbb{R}}
\newcommand{\IC}{\mathbb{C}}
\newcommand{\IZ}{\mathbb{Z}}

\newcommand{\tzeta}{\tilde\zeta}

\newcommand{\txi}{\tilde\xi}
\newcommand{\talp}{\tilde\alpha}

\def\bF{\bar F}
\def\bY{\bar Y}

\def\bV{ \bar V }

\def\bZ{\bar Z}
\def\bcY{\bar \cY}

\def\beuF{\bar \euF}

\def\btau{\bar \tau}

\def\bt{\bar t}

\def\ba{\bar a}

\def\bi{\bar \imath}

\def\bz{\bar z}

\def\hz{\hat z}

\def\hzeta{\hat \zeta}
\def\hgam{\hat \gamma}
\def\hp{\hat\p}

\def\ci#1{c^{[#1]}}
\def\cij#1{c^{[#1]}}

\def\txii#1{{\tilde\xi}^{[#1]}}
\def\ai#1{{\alpha}^{[#1]}}

\def\alpi#1{\alpha^{[#1]}}

\def\ellg#1{\ell_{#1}}

\def\Zg{Z_{\gamma}}

\def\bOm{\bar\Omega}

\def\scLn#1{\scL^{(#1)}}
\def\bscLn#1{\bar{\scL}^{(#1)}}
\def\scKn#1#2{\cK^{(#1)}_{\bfg,#2}}

\def\cJn#1{\cJ^{(#1)}}
\def\cIn#1{\cI^{(#1)}}
\def\cGn#1#2{\dIu^{(#1)#2}}
\def\tcGn#1#2{\dId^{(#1)}_{#2}}
\def\cLn#1#2{\cL^{(#1)}_{#2}}
\def\bcLn#1#2{\bar{\cL}^{(#1)}_{#2}}
\def\tccKn#1#2#3{\dIdd^{(#1)}_{\bfg,#2#3}}
\def\tcKn#1#2#3{\dIud^{(#1)#3}_{\bfg,#2}}
\def\cKn#1#2#3{\dIuu^{(#1)#2#3}_\bfg}
\def\ccKn#1#2#3{\dIdu^{(#1)#2}_{\bfg,#3}}

\def\XXint#1#2#3{{\setbox0=\hbox{$#1{#2#3}{\int}$}
\vcenter{\hbox{$#2#3$}}\kern-.5\wd0}}

\def\cij#1{c}
\def\ci#1{c}

\def\cCf{\cC}

\def\tleta{\tilde\eta}

\def\CY{\mathfrak{Y}}
\def\CYm{\mathfrak{\hat Y}}
\def\sCY{{\scriptstyle \CY}}

\def\hpert{h^{(0)}}
\def\zpert{z_0}
\def\bzpert{\bz_0}
\def\Fcl{F_{\rm cl}}

\def\frJ{\mathfrak{J}}
\def\frV{\mathfrak{V}}
\def\frD{\mathfrak{D}}

\def\frL{\mathfrak{L}}

\def\frS{\mathfrak{S}}
\def\frZ{\mathfrak{Z}}
\def\bfrL{\bar{\frL}}

\def\frLn#1#2{\frL^{(#1)}_{\bfg,#2}}
\def\bfrLn#1#2{\bar{\frL}^{(#1)}_{\bfg,#2}}

\def\euF{\EuScript{F}}

\def\rin{r_{\rm inst}}
\def\Win{W_{\rm inst}}
\def\bWin{\bar W_{\rm inst}}

\def\bfg{{\boldsymbol{\gamma}}}
\def\hbfg{\hat\bfg}

\def\dIu{\cI}
\def\dId{\cI}
\def\dIuu{\cI}
\def\dIud{\hat\cI}
\def\dIdu{\cI}
\def\dIdd{\cI}
\def\opI{\scJ}
\def\Prz{\mathscr{P}}

\title{Hypermultiplet metric and NS5-instantons}

\author{Sergei Alexandrov and Khalil Bendriss
\\
{\it Laboratoire Charles Coulomb (L2C), Universit\'e de Montpellier,
CNRS, \\ F-34095, Montpellier, France}\\

\vspace{2mm} {\tt e-mail:
\email{sergey.alexandrov@umontpellier.fr},
\email{khalil.bendriss@umontpellier.fr}
}

\vspace{-3mm}

}

\abstract{The metric on the hypermultiplet moduli space of Calabi-Yau compactifications of type II string theory
is known to receive D-brane and NS5-brane instanton corrections. We compute explicit expressions for these corrections
in the one-instanton approximation, but to all orders in the string coupling expansion around the instantons.
As a consistency check, we prove that in the case of one (universal) hypermultiplet,
the resulting metric fits the Przanowski description of self-dual Einstein spaces.
We also show that in the small string coupling limit the metric acquires a certain square structure,
consistently with expectations from the string amplitudes analysis. This result provides explicit predictions
for yet mysterious string amplitudes in the presence of NS5-branes.

}

\begin{document}

\textheight=24.3cm

\section{Introduction}

Instanton effects, although exponentially suppressed in the small coupling limit, play an extremely important role in string theory.
Typically, they are necessary ingredient for moduli stabilization, singularity resolution, correct analyticity properties,
but most importantly they give us a window into the realm of non-perturbative phenomena where
our understanding of string theory is still very poor.

The usual source of instantons in string theory are D-branes, either localized ones existing in type IIB (D(-1)-branes)
and in low-dimensional models (ZZ-branes), or extended ones wrapping non-trivial cycles of a compactification manifold.
In fact, until recently the actual calculation of D-instanton effects was possible only using various dualities
such as S-duality or the dual description of non-critical strings via matrix models, whereas their direct computation
from amplitudes of open strings with boundaries on D-instantons suffered from divergences.
In a series of remarkable works \cite{Sen:2020cef,Sen:2020eck,Sen:2021qdk} Ashoke Sen understood how all these divergences
can be regularized and eventually removed using string field theory. This made it possible not only to reproduce some
known results in critical and non-critical string theories
\cite{Sen:2021tpp,Sen:2021jbr,Alexandrov:2021shf,Alexandrov:2021dyl,Eniceicu:2022nay,Eniceicu:2022dru,Eniceicu:2022xvk},
but also to get new ones where constraints from supersymmetry and dualities are not powerful enough
to fix D-instanton contributions \cite{Alexandrov:2022mmy}.

However, there is another type of branes in string theory that gives rise to instanton effects.
These are NS5-branes of ten-dimensional string theory or M5-branes of M-theory.
They have six-dimensional world-volume and generate instantons if the target space has a six-dimensional
compact cycle which they can wrap.
To our knowledge, no direct calculation of string amplitudes responsible for such NS5-instantons has been done so far.

In such situation one may hope that one can get some insight from predictions of dualities,
as was the case for D-instantons until a few years ago.
One of the simplest setups where NS5-instantons appear is the compactification of type II string theory
on a Calabi-Yau (CY) threefold. In this case the effective action at the two-derivative level is
determined by the metric on the vector and hypermultiplet moduli spaces and the latter is affected by both
D-brane and NS5-brane instantons \cite{Becker:1995kb}.
The D-brane instantons are fairly well understood \cite{RoblesLlana:2006is,Alexandrov:2008gh,Alexandrov:2009zh} and,
at least in type IIA\footnote{Although the type IIB formulation can be obtained from type IIA by mirror symmetry,
to write the metric in terms of type IIB physical fields, one needs to know an instanton corrected mirror map,
whose classical version has been identified in \cite{Bohm:1999uk}. At this point it is known only
how to include corrections from D1-D(-1)-instantons \cite{Alexandrov:2009qq,Alexandrov:2012bu}
and from D3-instantons in the large volume limit \cite{Alexandrov:2012au,Alexandrov:2017qhn}.},
we have access to the D-instanton corrected hypermultiplet metric to arbitrary order in the instanton expansion
\cite{Alexandrov:2014sya,Alexandrov:2017mgi}.\footnote{See also \cite{Cortes:2021vdm,Cortes:2023oje}
for a proper mathematical treatment of some of these results.}
One could think that the knowledge of D-instantons should be sufficient to get NS5-instantons by
going to type IIB and applying S-duality since it maps D5 to NS5-branes. This idea has been pursued
in \cite{Alexandrov:2010ca,Alexandrov:2014mfa,Alexandrov:2014rca}, but the results are far from being conclusive.
There are two reasons for this.

First, the hypermultiplet moduli space $\cM_H$  is a quaternion-K\"ahler (QK) manifold \cite{Bagger:1983tt}.
Its quantum corrected metric is horribly complicated and the only way to apply constraints of S-duality to it
is to use a twistorial formalism for QK spaces
(see \cite{Alexandrov:2011va} for a review of applications of this formalism to CY compactifications).
This formalism allows to encode the metric into a set of holomorphic functions on
a $\CP$-bundle over $\cM_H$ known as the twistor space \cite{MR1327157}.
These functions, distant cousins of the holomorphic prepotential of special K\"ahler geometry,
are typically very simple. But the price to pay for this simplicity is a complicated procedure
of going from the data on the twistor space to the actual metric \cite{Alexandrov:2008nk}.
While it has been realized for D-instantons in \cite{Alexandrov:2014sya,Alexandrov:2017mgi},
no attempt has been done so far to apply it for NS5-brane instantons.

Second, the naive application of S-duality to the twistorial description of D-instantons
leads to a construction \cite{Alexandrov:2014mfa,Alexandrov:2014rca} which is mutually
inconsistent at the multi-instanton level. This indicates that there might be missing
S-duality invariant sectors of multi-instantons that are not captured by this approach.

Thus, at this point the twistorial construction of NS5-instantons is trustworthy only at one-instanton level,
but the corresponding corrections to the hypermultiplet metric, and hence to the effective action, have not been
derived yet even in this approximation. The goal of this paper is to fill in this gap.
Namely, we start from the twistorial description of NS5-instantons developed in \cite{Alexandrov:2010ca}
and, following the procedure described in \cite{Alexandrov:2008nk}, obtain a one-instanton contribution
generated by NS5-branes to the hypermultiplet metric.
The result can be found in \eqref{ds2final2}.

If the CY is rigid, i.e. it has $h^{2,1}=0$, there is only one hypermultiplet and $\cM_H$ is four-dimensional.
As was shown by Przanowski \cite{Przanowski:1984qq}, the metric on such QK manifolds is
described by one real function (Przanowski potential) satisfying a non-linear partial differential equation.
This provides a consistency check on our metric which must be compatible with this description.
We show that this is indeed the case and compute an explicit expression for the Przanowski potential
in the one-instanton approximation following from \eqref{ds2final2}.

Furthermore, evaluating the small string coupling limit of \eqref{ds2final2},
one can get a prediction for various string amplitudes in the presence of NS5-branes.
In fact, generalizing the reasoning of \cite{Alexandrov:2021shf} for D-instantons to the case of NS5-branes,
we show that at the leading order the one-instanton contribution must have the following square structure
\be
\de s^2_{\rm NS5}
\simeq\sum_\bfg C_\bfg\, e^{-S_\bfg}\( \cA_\bfg^2+\cB_\bfg \de S_\bfg\),
\label{sqaurestr}
\ee
where $\bfg$ is a charge vector labelling bound states of NS5 and D-branes, $S_\bfg$ is an instanton action,
$C_\bfg$ is a function of the moduli scaling as a power of $g_s$,
and $\cA_\bfg$, $\cB_\bfg$ are one-forms on $\cM_H$ such that the coefficients of $\cA_\bfg$ are given by
certain string amplitudes in an NS5-background, while $\cB_\bfg$ cannot be fixed by this analysis.
We show that the metric which we derived does exhibit the structure \eqref{sqaurestr}
with specific $C_\bfg$ and one-forms $\cA_\bfg$ and $\cB_\bfg$. This provides a prediction for three-point sphere
correlation functions where one of the vertex operators corresponds to a hypermultiplet scalar
and two others represent fermionic zero modes of the background NS5-brane.
While for generic background fields the prediction is somewhat involved, it drastically simplifies
in the limit of small RR fields (see \eqref{limit-onrforms}).
We hope that this prediction will help to understand how these amplitudes can be computed directly
using worldsheet techniques.

The organization of the paper is the following. In section \ref{sec-HM} we review
basic facts about the hypermultiplet moduli space and its twistorial description.
In particular, we formulate our starting point given by holomorphic functions
on the twistor space encoding D-brane and NS5-brane instantons.
In the next section we compute the metric following from this twistorial data in the one-instanton approximation.
In section \ref{sec-UHM} we consider the four-dimensional case of the universal hypermultiplet
and in section \ref{sec-smallgs} we compute the small $g_s$ limit of our metric.
Finally, section \ref{sec-concl} summarizes our results.
Several appendices contain details of the computations.

\section{Twistor description of the hypermultiplet moduli space}
\label{sec-HM}

\subsection{The moduli space}
\label{subsec-MH}

At low energies, type IIA string theory compactified on a CY threefold $\CY$ reduces to $N=2$ supergravity
coupled to $h^{1,1}$ vector multiplets and $h^{2,1}+1$ hypermultiplets.
These multiplets comprise $h^{1,1}$ complex and $4(h^{2,1}+1)$ real scalar fields
which parametrize the vector and hypermultiplet moduli spaces, $\cM_V$ and $\cM_H$, respectively.
Each of the moduli spaces carries a metric and can be thought as the target space of a non-linear sigma model
described by the kinetic terms for the corresponding scalars in the effective action,
which is completely fixed once the two metrics are known.
Furthermore, the local supersymmetry requires that $\cM_V$ is a projective special K\"ahler manifold, while
$\cM_H$ is a quaternion-K\"ahler manifold \cite{Bagger:1983tt}.

Physically, the scalar fields parametrizing $\cM_V$ represent complexified K\"ahler moduli of $\CY$,
whereas $\cM_H$ comprises four different types of fields:
\begin{itemize}
\item
complex structure deformations of $\CY$, denoted by $z^a$ ($a=1,\dots,h^{2,1}$);

\item
RR fields $\zeta^\Lambda,\tzeta_\Lambda$ ($\Lambda=0,\dots,h^{2,1}$)
arising as period integrals of the RR 3-form of type IIA string theory over a symplectic basis of cycles in $H_3(\CY,\IZ)$;

\item
four-dimensional dilaton\footnote{Throughout the paper we use the name `dilaton' for its exponential given by the field $r$.}
$r\equiv e^{-2\phi_{(4)}} \sim g_s^{-2}$;

\item
and NS axion $\sigma$ which is dual to the $B$-field in four dimensions.
\end{itemize}
Note that the subspace $\cM_C\subset\cM_H$ parametrized by $z^a$ carries a special K\"ahler geometry similarly to $\cM_V$.
This is because it is identified with $\cM_V$ in the mirror type IIB formulation where
the roles of K\"ahler and complex structure moduli are exchanged. As a result, this subspace is characterized
by a prepotential $F(X)$, a holomorphic homogeneous function of degree 2, where the homogeneous coordinates $X^\Lambda$
are related to the moduli by $z^a=X^a/X^0$. In particular, it defines the K\"ahler potential as
\be
\cK=-\log K,
\qquad
K  = -2\Im (\bz^\Lambda F_\Lambda(z)),
\label{Kahler}
\ee
where $F_\Lambda=\p_{X^\Lambda}F$ and we defined $z^\Lambda=(1,z^a)$.

In fact, the prepotential also determines the full perturbative metric on $\cM_H$
which has the following explicit expression \cite{Alexandrov:2007ec}
(obtained on the basis of earlier works \cite{Antoniadis:1997eg,Gunther:1998sc,Antoniadis:2003sw,Robles-Llana:2006ez})
\be
\begin{split}
\de s^2_{\rm pert} =&\, \frac{r+2c}{r^2(r+c)}\, \de r^2
-\frac{1}{r} \(N^{\Lambda\Sigma} - \frac{2(r+c)}{rK}\, z^\Lambda \bz^\Sigma\) \(\de \tzeta_\Lambda - F_{\Lambda\Lambda'} \de \zeta^{\Lambda'}\)
\(\de\tzeta_\Sigma -\bF_{\Sigma\Sigma'} \de \zeta^{\Sigma'}\)
\\
&\,
+\frac{r+c}{16 r^2(r+2c)} \(\de\sigma + \tzeta_\Lambda \de \zeta^\Lambda - \zeta^\Lambda \de\tzeta_\Lambda + 8c \cA_K\)^2
+ \frac{4(r+c)}{r}\, \cK_{a\bar{b}} \de z^a \de \bz^b.
\end{split}
\label{1lmetric}
\ee
Here we denoted $N_{\Lambda\Sigma} = -2\Im F_{\Lambda\Sigma}$, the matrix $N^{\Lambda\Sigma}$ is its inverse,
and
\be
\cA_K =  \frac{\I}{2}\(\cK_a\de z^a-\cK_{\bar a}\de \bz^a\)=\Im\p\log K
\label{kalcon}
\ee
is the K\"ahler connection on $\cM_C$. Finally, the numerical parameter $c$
is given by
\be
c = -\frac{\chi_\sCY}{192\pi}\, ,
\ee
where $\chi_\sCY$ is the Euler characteristic of $\CY$,
and encodes the one-loop $g_s$-correction to the classical metric.
If it is set to zero, the metric \eqref{1lmetric} reduces to the so-called c-map
\cite{Cecotti:1989qn,Ferrara:1989ik} which gives a canonical
construction of a QK manifold as a bundle over a special K\"ahler base.

Although there are no perturbative $g_s$-corrections beyond one loop \cite{Robles-Llana:2006ez},
the metric \eqref{1lmetric} receives non-perturbative corrections from D2-branes wrapping 3-dimensional cycles on $\CY$
and NS5-branes wrapping the whole CY. They scale in the small string coupling limit roughly as
$e^{-1/g_s}$ and $e^{-1/g_s^2}$, respectively.
As explained in the introduction, the most efficient way to incorporate them is
to use the twistorial description of $\cM_H$.
In the next subsection we recall some elements of this description that will be necessary for our purposes
(see \cite{Alexandrov:2011va} for a more comprehensive review).

Before that let us say a few words about symmetries of $\cM_H$.
First, the hypermultiplet metric \eqref{1lmetric} carries an action of the symplectic group $Sp(2h^{2,1}+2,\IZ)$,
It leaves $r$ and $\sigma$ invariant and transforms $(X^\Lambda,F_\Lambda)$ and $(\zeta^\Lambda,\tzeta_\Lambda)$
as vectors. However, since generic symplectic transformation affects the prepotential $F$, it is not a true isometry of $\cM_H$.
Only a subgroup of $Sp(2h^{2,1}+2,\IZ)$ which is realized as monodromies around singularities of the complex structure moduli space
is a true isometry.
The symplectic invariance can be seen as a characteristic feature of the type IIA formulation
and is expected to hold at the non-perturbative level.

The metric \eqref{1lmetric} is also invariant under Peccei-Quinn symmetries acting by shifts on the RR fields and the NS axion
\be
T_{\eta^\Lambda,\tleta_\Lambda,\kappa}\ : \
(\zeta^\Lambda,\tzeta_\Lambda,\sigma)\mapsto
(\zeta^\Lambda+\eta^\Lambda,\tzeta_\Lambda+\tleta_\Lambda,\sigma+2\kappa-\tleta_\Lambda\zeta^\Lambda+\eta_\Lambda\tzeta_\Lambda).
\label{Heis}
\ee
At the perturbative level, the parameters $(\eta^\Lambda,\tleta_\Lambda,\kappa)$ can take any real value, whereas
instanton corrections break these isometries to a discrete subgroup with $(\eta^\Lambda,\tleta_\Lambda,\kappa)\in\IZ^{2h^{2,1}+3}$.
In particular, D-instantons break continuous shifts of the RR fields, but leave the invariance along $\sigma$, while
NS5-instantons break them all. The fact that the transformations \eqref{Heis} form the non-commutative
Heisenberg algebra plays an important role for description of NS5-instantons
(see, e.g., \cite{Witten:1996hc,Dijkgraaf:2002ac,Pioline:2009qt,Bao:2009fg}).

Finally, mirror symmetry implies that $\cM_H$ in type IIA compactified on $\CY$ is identical to
the same moduli space in type IIB compactified on a mirror CY $\CYm$. Furthermore,
in this mirror type IIB formulation, $\cM_H$ must carry an isometric action of the S-duality group $SL(2,\IZ)$.
A consistent realization of all expected discrete isometries on type IIB fields at full non-perturbative level has been found
in \cite{Alexandrov:2014rca}. By mirror symmetry, it can be mapped to type IIA so we assume that this problem is solved
and do not discuss it anymore in this paper.

\subsection{The twistor space}
\label{subsec-twistor}

The hypermultiplet moduli space is an example of QK manifold which is defined as $4n$ real dimensional Riemannian manifold $\cM$
with holonomy group $Sp(n)\times SU(2)$. It carries a quaternionic structure given by a triplet of almost complex structures
$J^i$, $i=1,2,3$, satisfying the algebra of quaternions. A generic almost complex structure is a normalized linear
combination of the triplet and parametrized by a point on $\CP$. This fact can be used to construct a canonical $\CP$-bundle
over $\cM$, known as its twistor space $\cZ_\cM$ \cite{MR1327157}.
It turns out that $\cZ_\cM$ is K\"ahler and, most importantly, carries a holomorphic contact structure
defined by the kernel of a canonical (1,0)-form
\be
D t = \de t + p^+ - \I t p^3 + t^2 p^- ,
\qquad
p^\pm=-\hf\(p^1\mp \I p^2\),
\label{canform}
\ee
where $t$ is a complex coordinate parametrizing the $\CP$ fiber and $\vec p$ is
the $SU(2)$ part of the Levi-Civita connection on $\cM$.
It is however more convenient to work in terms of a {\it holomorphic} one-form
\be
\cX = \frac{4}{\I t} \,e^{\Phi(t)}\, D t
\label{relcontform}
\ee
defined up to multiplication by a non-vanishing holomorphic function. The function $\Phi$ determining the rescaling
coefficient, which makes the (1,0)-form holomorphic,
is called contact potential \cite{Alexandrov:2008nk} and will play an important role in what follows.

A crucial feature of the contact one-form is that locally, by a proper choice of coordinates,
it can always be trivialized to the canonical form
\be
\cX = -\hf\(\de \talp + \txi_\Lambda\de\xi^\Lambda -\xi^\Lambda \de \txi_\Lambda\),
\label{contform}
\ee
where $(\xi^\Lambda,\txi_\Lambda,\talp)$ is a set of {\it holomorphic Darboux coordinates}.
These coordinates play a central role in the twistorial construction because
their knowledge as functions of coordinates $(\vph^m,t)$ on the base and on the fiber
of the twistor bundle allows to get the metric on $\cM$:
by combining \eqref{relcontform} and \eqref{contform}, one can find the $SU(2)$ connection $\vec p$
which can then be used to get the almost complex structure $J^3$ and
compute the triplet of quaternionic two-forms $\vec \omega$
\be
\vec \omega=-4\({\rm d} \vec p + \frac{1}{2}\,\vec p \times\vec p\),
\label{su2curvature}
\ee
and then the metric follows from $g(X,Y) = \omega^3(X,J^3 Y)$.\footnote{The overall normalization of $\vec \omega$ in \eqref{su2curvature}
is related to the value of the Ricci scalar of the metric on $\cM$.
We have chosen it to be compatible with the perturbative metric \eqref{1lmetric}.}
More details on this procedure are presented in appendix \ref{apap-metricgen}.

The Darboux coordinates encoding the perturbative metric \eqref{1lmetric} on $\cM_H$
have been found in \cite{Neitzke:2007ke,Alexandrov:2008nk} and are given by
\be
\label{exline}
\begin{split}
\xi_{\rm pert}^\Lambda =&\, \zeta^\Lambda + \cR \left(
t^{-1} z^\Lambda - t \, \bz^\Lambda\right),
\\
\txi^{\rm pert}_\Lambda =&\,
\tzeta_\Lambda
+\cR \left( t^{-1} F_\Lambda -t \, \bF_\Lambda \right) ,
\\
\talp^{\rm pert}=&\,\sigma+\cR\,\(t^{-1} W-t \bar W\)-8\I c \log t,
\end{split}
\ee
where
\be
\label{defWnonThkl}
W = F_\Lambda \zeta^\Lambda - z^\Lambda \tzeta_\Lambda
\ee
and the coordinate $\cR$ is related to the K\"ahler potential \eqref{Kahler} and the contact potential,
which in this case is $t$-independent and identified with the dilaton
\be
r\equiv e^{\Phi_{\rm pert}}=\frac{1}{4}\, \cR^2 K-c.
\label{pert-r}
\ee

We are interested in the deformations of the perturbative moduli space generated by D-brane and NS5-brane instanton corrections
and preserving the QK structure. Since we are interested only in the one-instanton approximation,
we can restrict ourselves to {\it linear} deformations and ignore all non-linear effects.
In this approximation, a generic deformation can be encoded in the twistor formalism into a set of contours $C_i$ on $\CP$
and an associated set of holomorphic functions $H_i(\xi,\txi,\talp)$ \cite{Alexandrov:2008nk}.
They lead to the following modification of the Darboux coordinates \eqref{exline}:
\be
\begin{split}
&
\xi^\Lambda  = \xi_{\rm pert}^\Lambda
+\frac{1}{4\pi \I} \sum_i \opI_i\[\hp_{\txi_\Lambda }H_i\] ,
\qquad\qquad
\txi_\Lambda =   \txi^{\rm pert}_\Lambda
-\frac{1}{4\pi \I}   \sum_i \opI_i\[\hp_{\xi^\Lambda } H_i\],
\\
&
\talp =  \talp^{\rm pert}
+\cR\(t^{-1}\Win-t\bWin\)
+\frac{1}{4\pi \I}   \sum_i \opI_i\[\( 2- \xi^\Lambda \hp_{\xi^\Lambda}-\txi_\Lambda \hp_{\txi_\Lambda}\)H_i\],
\end{split}
\label{genDC}
\ee
where we introduced a linear integral transform acting on holomorphic functions on the twistor space
\be
\opI_i[H]=\int_{C_i} \frac{\de t'}{t'} \,
\frac{t' + t}{t' - t}\, H\(\xi_{\rm pert}(t'), \txi^{\rm pert}(t'), \talp^{\rm pert}(t')\)
\label{def-frI}
\ee
as well as
\be
\hp_{\xi^\Lambda }=\p_{\xi^\Lambda }-\txi_\Lambda\p_{\talp}\, ,
\qquad
\hp_{\txi_\Lambda }=\p_{\txi_\Lambda }+\xi^\Lambda \p_{\talp }\, ,
\label{def-hatder}
\ee
\be
\Win=\frac{1}{4\pi \I}\sum_i \int_{C_i} \frac{\de t'}{t'}
\( z^\Lambda\,\hp_{\xi^\Lambda }+F_\Lambda \,\hp_{\txi_\Lambda} \)H_i^{\rm pert}\, .
\ee
To get a real metric from \eqref{genDC}, one should impose an additional condition that
the set $\{ C_i\}$ is invariant under the antipodal map $\varsigma[t]=-1/\bt$, while the set $\{H_i\}$
is invariant under the combination of $\varsigma$ with complex conjugation, i.e. for each $i$ there is $\bi$
such that
\be
\varsigma[C_i]=C_{\bi},
\qquad
\overline{\varsigma[H_i]}=H_{\bi}.
\label{reality}
\ee

It is clear that the deformation makes the Darboux coordinates to be multi-valued functions of $t$ with jumps
across $C_i$ determined by $H_i$. Therefore, $H_i$ have the meaning of {\it transition functions}.
Alternatively, one can think that the contours $C_i$ separate different patches on $\CP$ each having its own set of
holomorphic Darboux coordinates. This is consistent with the fact that the contact one-form \eqref{contform}
and hence the Darboux coordinates trivializing it are defined only locally.

To complete the twistorial description of linear deformations, we give also a formula
for a modification of the contact potential $\Phi$ which is given by
\be
\Phi(t)=\phi+\frac{1}{2\pi \I} \sum_i \opI_i\[ \p_{\talp}H_i\],
\label{eqchip}
\ee
where the $t$-independent part reads as
\be
e^\phi=\frac{1}{4}\,\cR^2 K-c -\frac{\cR}{16\pi} \sum_i\int_{C_i}\frac{\de t}{t}
\[\(t^{-1} z^{\Lambda}-t \bz^{\Lambda} \)\hp_{\xi^\Lambda }
+\(t^{-1} F_{\Lambda}-t \bF_{\Lambda} \)\hp_{\txi_\Lambda }\]H_i^{\rm pert}\, .
\label{contpotconst}
\ee

\subsection{Twistorial construction of instantons}
\label{subsec-insttwist}

In this subsection we provide the twistor data that
incorporate D-brane and NS5-brane instanton corrections.
As explained above, the data consist of a set of contours $C_i$ on $\CP$
and an associated set of holomorphic transition functions $H_i$.

\subsubsection{D-instantons}

D-instantons have been incorporated into the twistorial description of $\cM_H$ at linear order in \cite{Alexandrov:2008gh}
and to all orders in the instanton expansion in \cite{Alexandrov:2009zh}.
Here we present the first simplified version which is sufficient at one-instanton level and fits the framework described above.

Let us recall that each D-instanton is characterized by a charge vector $\gamma=(p^\Lambda, q_\Lambda)$.
In type IIA, it is integer valued and characterizes the 3-cycle wrapped by D2-brane generating the instanton,
in the same basis of $H_3(\CY,\IZ)$ that is used to define RR fields $(\zeta^\Lambda,\tzeta_\Lambda)$.
Given the charge, we introduce the central charge function
\be
\label{defZ}
Z_\gamma(z) = q_\Lambda z^\Lambda- p^\Lambda F_\Lambda(z),
\ee
and the generalized Donaldson-Thomas (DT) invariant $\Omega_\gamma$. Although a proper mathematical definition
of this topological invariant is quite involved \cite{Joyce:2008pc}, from the physical viewpoint it just counts
the number of BPS instantons of a given charge. In the following we will mainly use its rational version
\be
\label{defntilde}
\bOm_\gamma = \sum_{d|\gamma}  \frac{1}{d^2}\,  \Omega_{\gamma/d} ,
\ee
which takes into account multi-covering effects and
allows to simplify many equations being more suitable for implementing S-duality \cite{Manschot:2010xp,Alexandrov:2012au}.
An important property of DT invariants is that $\Omega_{-\gamma}=\Omega_\gamma$.

Finally, we define the so-called BPS ray
\be
\ellg{\gamma}= \{ t :\,  Z_\gamma(z)/t \in \I\IR^{-} \},
\label{rays}
\ee
which joins the north and south poles of $\CP$ along the direction determined by the phase of the central charge,
and the following transition function assigned to $\ell_\gamma$
\be
H_\gamma
=\frac{\sigma_\gamma\,\bOm_\gamma}{4\pi^2}\, e^{-2\pi \I (q_\Lambda \xi^\Lambda-p^\Lambda\txi_\Lambda)} ,
\label{prepH}
\ee
where $\sigma_\gamma$ is the so-called quadratic refinement. This is a sign factor that must satisfy
$
\sigma_{\gamma+\gamma'} =
(-1)^{\langle \gamma, \gamma' \rangle}\, \sigma_\gamma\, \sigma_{\gamma'}$,
where $\langle\gamma,\gamma'\rangle=q_\Lambda p'^\Lambda-q'_\Lambda p^\Lambda$ is the skew-symmetric product of charges.
In the following it is chosen to be $\sigma_\gamma=(-1)^{q_\Lambda p^\Lambda}$.\footnote{How this choice is reconciled
with symplectic invariance is explained in \cite[\$2.3]{Alexandrov:2014rca}.}
The set of all $(\ell_\gamma,H_\gamma)$ for which DT invariants are non-vanishing comprise the twistor data of D-instantons.
They affect the Darboux coordinates according to equations \eqref{genDC} and the resulting corrections to the metric
have the form of D-instanton corrections. Their explicit form will be obtained below.

Note that our one-instanton approximation corresponds to keeping only terms linear in $\bOm_\gamma$,
while we allow for D-instantons of different (in particular, proportional) charges.
Thus, it is not about extracting the dominant instanton contribution, but rather the linear response of the metric
to the change of the contact structure by $(\ell_\gamma,H_\gamma)$.
This approach allows us to get results independent of particular values of DT invariants and
to keep track of the charge dependence in the resulting instanton corrections.

\subsubsection{NS5-instantons}

The twistor data incorporating NS5-instantons in the one-instanton approximation (as it was defined in the previous subsection)
have been found in \cite{Alexandrov:2010ca} by translating the above construction of D-instantons
to the mirror type IIB formulation and applying S-duality to D-instantons with a non-vanishing D5-brane charge.
The duality was applied at the level of the twistor space where it acts by a holomorphic transformation preserving
the contact structure. In particular, its action on the fiber coordinate $t$ and the Darboux coordinates
$(\xi^\Lambda,\txi_\Lambda,\talp)$ is well known \cite{Alexandrov:2008nk,Alexandrov:2012bu}
and therefore allows to get the contours and transition functions incorporating NS5-instantons as images of $(\ell_\gamma, H_\gamma)$
under this action.

We would like to borrow these results to derive the one-instanton corrected metric in type IIA.
However, by construction, the described procedure leads to a twistor space adapted to the type IIB formulation.
In particular, it is manifestly invariant under $SL(2,\IZ)$ duality group, rather than under symplectic transformations,
and the resulting transition functions are expressed in terms
of topological data characterising the K\"ahler moduli space of the mirror CY.
So one may ask whether one can use the transition functions of the type IIB formulation in type IIA.
We claim that it is possible because geometrically $\cM_H$ and its twistor space are the same in the two formulations
and differ only by the choice of coordinates (physical fields) used to express the metric.

In fact, any contact structure can also be encoded using different sets of contours and transition functions.
For example, in our context they can be adapted to symmetries of type IIA and type IIB, respectively.
However, the two sets must be related by a contact transformation. An explicit example is provided
by the case of D1-D(-1)-instantons where such transformation has been explicitly found
\cite{Alexandrov:2009qq,Alexandrov:2012bu}.
The contact transformation can affect other sets of transition functions \cite{Alexandrov:2014rca},
but only at multi-instanton level. In the one-instanton approximation, different sets do not interfere.
Therefore, in this approximation, it should not matter which formulation of the contact structure one exploits
to get the metric on the moduli space.

Thus, we take the twistor data encoding NS5-instantons obtained in \cite{Alexandrov:2010ca},
translate them back into type IIA language, and apply as deformations of the perturbative twistor space.
The price to pay for using type IIB twistor data in type IIA is the absence of manifest symplectic invariance.
We also believe that the complicated structure of the resulting metric is partially a consequence of this hybrid approach
and there should exist a genuine type IIA formulation of NS5-instantons.
However, in the absence of such formulation, we have to proceed as just described, but we hope that
our results can shed light on this and other issues related to the geometry of $\cM_H$.

After these preliminary comments, let us describe the twistor data for NS5-instantons
after they have been translated (partially) to type IIA variables in appendix \ref{ap-HNS5}.
To this end, let us introduce an integer valued charge vector
$\bfg=(k,p,\hgam)$ with $k\ne 0$ and $\hgam=(p^a,q_a,q_0)$.
Here $k$ denotes NS5-brane charge, while the other components are related to bound D-branes.
In particular, the standard D-brane charge vector is obtained as
$\gamma=(p^0,\hgam)$ where $p^0=\gcd(k,p)$.
On the type IIB side, $p^0$ is D5-brane charge, while $\hgam$ encodes D3-D1-D(-1)-charges.

Given the charge $\bfg$, we define the contour $\ell_{\bfg}$ as a half-circle\footnote{More precisely,
$\ell_{\bfg}$ is the image of the BPS ray $\ell_\gamma$ under
(the inverse of) $SL(2,\IZ)$ transformation $g_{k,p}$ \eqref{Sdualde}: $\ell_{\bfg}=g_{k,p}^{-1}\cdot \ell_\gamma$.
In particular, the transformation preserves the ordering of the original BPS rays.
}
stretching between the two zeros of $\xi^0_{\rm pert}(t)-p/k$
and the associated transition function
\be
\label{5pqZ2}
H_{\bfg}=\frac{\bOm_\gamma}{4\pi^2}\,e^{-\pi \I k \(\talp+(\xi^\Lambda-2n^\Lambda) \txi_\Lambda\)}
\Psi_{\hbfg}(\xi-n) ,
\ee
where
$\hbfg$ is a reduced charge
\be
\hbfg=\bfg \mod \(0,k\eps^0,\delta[\gamma,k\eps^a/p^0]\),
\qquad
\eps^\Lambda\in\IZ
\label{def-redgam}
\ee
with $\delta[\gamma,\eps^a]$ being the spectral flow transformation \eqref{def-spflowq},
and
\be
\Psi_{\hbfg}(\xi)=- \,\frac{k\xi^0}{p^0} \, \exp\[2\pi\I\(
k F(\xi)
+\frac{m_a\xi^a +Q}{\xi^0}+m_0
\) \],
\qquad k>0,
\label{Psibfg}
\ee
while for negative $k$ the holomorphic prepotential $F$ should be replaced by its complex conjugate $\bF$.
To write these definitions, we introduced various rational charges
\be
\begin{split}
& \quad\quad
n^0=p/k,
\qquad
n^a=p^a/k,
\qquad
Q=(p^0)^2\hat q_0/k^2,
\\
m_0=&\,a p^0 q'_0/k- c_{2,a}p^a \varepsilon(g_{k,p})-\frac{k}{2}\, A_{\Lambda\Sigma}n^\Lambda n^\Sigma+\hf\, q_\Lambda p^\Lambda,
\qquad
m_a=p^0\hat q_a/k.
\end{split}
\label{defma0}
\ee
The notations $\hat q_\Lambda$, $q'_a$, $A_{\Lambda\Sigma}$, $c_{2,a}$, $a$ and $\varepsilon(g_{k,p})$ are explained in appendix \ref{ap-HNS5}.

In fact, most of our results hold for arbitrary function $\Psi_{\hbfg}$ (provided it ensures
convergence of integrals along $\ell_{\bfg}$). Its concrete form will be important only in deriving
the small string coupling approximation in section \ref{sec-smallgs}. On the other hand, the form of \eqref{5pqZ2} is dictated by
the Heisenberg symmetry \eqref{Heis} which acts on the Darboux coordinates $(\xi^\Lambda,\txi_\Lambda,\talp)$
in the same way as on the real coordinates $(\zeta^\Lambda,\tzeta_\Lambda, \sigma)$. Since it is non-commutative, only
the $\kappa$ and $\tleta$-shifts are realized in a simple way as symmetries of the transition functions.
In contrast, the $\eta$-shift maps different transition functions to each other, which is ensured by the fact
that $\Psi_{\hbfg}$ depends only on the reduced charge \eqref{def-redgam}
and the combination $\xi^\Lambda-n^\Lambda$.\footnote{In fact, in our approximation
where we ignore wall-crossing phenomena, the DT invariant $\bOm_\gamma$ does not change under the monodromies \eqref{monmat}
and therefore it can also be thought as a function of the reduced charge $\hbfg$
and combined with $\Psi_{\hbfg}$.}
The (hidden) symplectic invariance ensures that it should be possible to rewrite the construction of NS5-instantons
in other ``frames" where a different set of shifts of the RR fields is trivialized.
However, a map between different frames is expected to be non-trivial and
to involve an integral transform (see, e.g., \cite{Alexandrov:2015xir}),
similarly to a change between coordinate and momentum representations in quantum mechanics.
In fact, for $k=1$ it was shown \cite{Alexandrov:2010ca} that the sum over electric charges $q_\Lambda$ of
$\Psi_{\hbfg}$ gives the topological string partition function
in the real polarization and thus indeed behaves as a wave function.

\section{Instanton corrected metric}
\label{sec-metric}

In this section we derive the one-instanton corrected metric on $\cM_H$ from
the twistorial formulation of D-brane and NS5-brane instantons given in section \ref{subsec-insttwist}.

Before we start, let us make a comment about the choice of coordinates.
As is clear already from the twistorial description of the perturbative moduli space, the natural variable appearing
in parametrization of the Darboux coordinates is $\cR$ rather than the dilaton field $r$.
At the perturbative level, they are related by a relatively simple relation \eqref{pert-r}.
However, non-perturbative corrections can further modify it.
Given that the perturbative expression for $r$ \eqref{pert-r} coincides with the contact potential,
it is natural to extend this identification also to the non-perturbative level.
This is what was done in \cite{Alexandrov:2008gh,Alexandrov:2011va,Alexandrov:2014sya},
where D-instantons have been incorporated, by setting
\be
r=e^\phi,
\label{rphi}
\ee
where $e^\phi$ is given by \eqref{contpotconst} with $H_i$ being the D-instanton transition functions $H_\gamma$ \eqref{prepH}.
However, in the presence of NS5-brane instantons the situation is not so clear. First, the full contact potential \eqref{eqchip}
becomes $t$-dependent due to the $\talp$-dependence of the transition functions \eqref{5pqZ2} and therefore it cannot be identified with $r$.
This still leaves us the possibility to equate $r$ to the real $t$-independent part of the contact potential as in \eqref{rphi}.
But this cannot be true because the dilaton is expected to have a simple transformation under S-duality (it must transform as
a modular form of weight $(-\hf,-\hf)$ \cite{Alexandrov:2008gh}), whereas NS5-instantons spoil the modular
transformation of $e^\phi$. In \cite{Alexandrov:2013mha} a modular invariant modification of the contact potential has been constructed
which thus represents a natural candidate for the definition of the dilaton field at the non-perturbative level.
However, if we used that definition, our results would have become even more complicated.
Since we work in type IIA where S-duality is not manifest anyway, we take the simpler option\footnote{Another option
is to take $\cR$ as an independent coordinate and do not introduce the four-dimensional dilaton at all.
In particular, this makes sense because $\cR$ is closely related to the ten-dimensional string coupling.
This option was accepted in \cite{Cortes:2021vdm} and led to a somewhat nicer expression for the D-instanton corrected metric.
However, given that the perturbative metric in type IIA is more natural in terms of $r$, we do not follow this possibility.}
\eqref{rphi} and do not claim that $r$ is exactly the physical dilaton
(although for simplicity we continue calling it `dilaton field').
We consider the identification \eqref{rphi} just as a possible choice of coordinates.
Although different choices lead to different expressions for the instanton corrections to the metric,
importantly, they do not affect predictions for string amplitudes obtained in the small string coupling limit
and given by the one-form $\cA_\bfg$ in \eqref{sqaurestr} since any coordinate changes leaving the perturbative metric intact
contribute only to the second term in that expression.

Thus, our goal is to compute the QK metric induced by the deformations \eqref{prepH} and \eqref{5pqZ2}
in the coordinates described in section \ref{subsec-MH} where $r$ is identified with \eqref{contpotconst}.
The starting point is the Darboux coordinates \eqref{genDC} where the index $i$ runs over the two sets of charges, $\gamma$ and $\bfg$.
The procedure to get the metric from these data is described in detail in appendix \ref{apap-metricgen}.
Since it is quite technical and not illuminating, we relegate all details of the calculations to appendix \ref{ap-detailsmetric}
and present only the final result.

To this end, we have to introduce several notations. First, we define the one-forms
\be
\begin{split}
\frZ_\Lambda=&\,\de\tzeta_\Lambda-F_{\Lambda\Sigma}\de\zeta^\Sigma,
\qquad
\frJ=z^\Lambda \frZ_\Lambda=z^\Lambda\de\tzeta_\Lambda-F_\Lambda\de\zeta^\Sigma,
\\
&\,\quad
\frS=\frac{1}{4}\(\de \sigma + \tzeta_\Lambda \de \zeta^\Lambda - \zeta^\Lambda \de \tzeta_\Lambda+8c\cA_K \),
\end{split}
\label{def-pertforms}
\ee
which arise naturally already in the perturbative metric \eqref{1lmetric}.
Besides, we also define a one-form labelled by D-instanton charge
\be
\cCf_\gamma= N^{\Lambda\Sigma}\(q_\Lambda-\Re F_{\Lambda\Xi}p^\Xi\)\(\de\tzeta_\Sigma-\Re F_{\Sigma\Theta}\de\zeta^\Theta\)
+\frac14\, N_{\Lambda\Sigma}\,p^\Lambda\,\de\zeta^\Sigma.
\label{connC}
\ee
Finally, we will use various functions labelled by charges $\gamma$ and $\bfg$ which are defined in appendix \ref{ap-fun}
as expansion coefficients of the integral transform \eqref{def-frI} of the transition functions and its derivatives.
While for D-instantons, using integration by parts,
all relevant quantities can be expressed only through three such functions, $\cJn{n}_\gamma$ with $n=0,\pm1$,
for NS5-instantons this does not seem to be possible and we have to deal with many different functions (see \$\ref{ap-NS5fun}).
Using all these definitions, the one-instanton corrected metric on $\cM_H$ is found to be
\be
\de s^2  =\de s^2_{\rm pert}
+\frac{\cR}{8 \pi^2 r}\sum_{\gamma}\sigma_\gamma \bOm_\gamma \, \frD_\gamma
+\frac{\cR}{8\pi^2r }\sum_{\bfg} \bOm_\gamma \,k\, \frV_{\bfg},
\label{ds2final2}
\ee
where $\frD_\gamma$ and $\frV_{\bfg}$ encode the D-brane and NS5-brane instanton contributions, respectively,
They are given by
\bea
\frD_\bfg &=&
\frac{\pi}{\cR}\cJn{0}_\gamma \[\cR^2 |Z_\gamma|^2 \(\(\frac{\de r}{r+c}+\de\log\frac{|Z_\gamma|^2}{K}\)^2
+\(\frac{\frS}{r+2c}+2\Im\p\log\frac{Z_\gamma}{K}\)^2\)+\de\Theta_\gamma^2-4\cC_\gamma^2
\]
\nn\\
&+&
\( Z_\gamma\cJn{1}_\gamma+\bZ_\gamma\cJn{-1}_\gamma\)\Biggl[
\frac{\I\frS^2}{4(r+2c)^2}-\frac{\I(r+c)\,\cA_K\frS}{r(r+2c)}
-\frac{\I\de r\,\de\log K}{4(r+c)}-\frac{\I |\frJ|^2}{2r K}
-\I\cK_{ab}\de z^a \de \bz^b
\nn\\
&&
+2\pi\cC_\gamma\(\frac{\frS}{r+2c}-2\,\cA_K\)
\Biggr]
+\(\frac{r+c}{r(r+2c)}\,\frS-4\pi \I\cC_\gamma \) \Bigl(\cJn{1}_{\gamma}\de Z_\gamma -\cJn{-1}_{\gamma}\de\bZ_\gamma\Bigr)
\nn\\
&+&
\frac{\I\de r}{2(r+c)}\, \de \Bigl(Z_\gamma \cJn{1}_{\gamma}+\bZ_\gamma \cJn{-1}_{\gamma}\Bigr)
- \frac{\I\cR}{r}\,\Im(\bZ_\gamma\,\frJ)\,\de \cJn{0}_{\gamma}.
\label{frDtot2}
\eea
and
\bea
\frV_\bfg &=&
\(\scLn{1}_\bfg+\bscLn{-1}_\bfg\) \Biggl[
\frac{\I}{8}\(\frac{(\de r)^2}{(r+c)^2}-\frac{\frS^2}{(r+2c)^2}\)
+\frac{\I}{2r}\,\cA_K\frS
+\frac{\I|\frJ|^2}{2r K}
+\frac{\I}{2}\,\frac{|\p K|^2}{K^2}+\I\cK_{ab}\de z^a \de \bz^b
\Biggr]
\nn\\
&-&
\frac{\I}{2}\(\cLn{1}{\bfg,\Lambda} \de z^\Lambda+\bcLn{-1}{\bfg,\Lambda}\de\bz^\Lambda\) \de \log K
-\(\cLn{1}{\bfg,\Lambda} \de z^\Lambda-\bcLn{-1}{\bfg,\Lambda}\de\bz^\Lambda\)\(\frac{\frS}{2r}+\cA_K\)
\nn\\
&-&
\frac{2(r+c)}{r(r+2c)}\,\frS\,\bigg[\frac{\I}{\cR}\, \cIn{0}_\bfg\(\frac{c\de r}{r+c}+r\de\log K\)
+\Re( \cIn{1}_\bfg\,\frJ)
\bigg]
\nn\\
&-&
2\pi k \Biggl\{
\frac{2}{\cR} \,\cIn{0}_\bfg\biggl(\frac{(r+2c)^2}{(r+c)^2}\,(\de r)^2
-\frS^2+\frac{2r(r+2c)}{r+c}\, \de r\, \de\log K
-4r\cA_K\frS-\frac{4(r+c)}{K}\,|\frJ|^2
\nn\\
&&\quad
+\frac{r^2(\de K)^2}{K^2}-4r^2\cA_K^2
\biggr)
+\cR\Im\(\(\scLn{2}_\bfg-\bscLn{0}_\bfg\)\frJ\) \(\frac{\frS}{r+2c}-2\cA_K\)
\nn\\
&&
+ 2\(\scLn{1}_\bfg+\bscLn{-1}_\bfg\)
\(\frac{\de r}{r+c}\(\frS-2c\cA_K\) -\de\log K\( \frac{c\frS}{r+2c}+2r\cA_K\)\)
\nn\\
&&
+4\(\frac{r+2c}{r+c}\, \de r+r\de\log K\)\biggl[\Im(\cIn{1}_\bfg\frJ)
-\cR^{-1}\Re\(N^{\Lambda\Sigma}\bcLn{0}{\bfg,\Lambda}\,\frZ_\Sigma\)
+\Im\(\cLn{1}{\bfg,\Lambda}\de z^\Lambda\)\biggr]
\nn\\
&&
+4\Re\frJ
\[\Im\( N^{\Lambda\Sigma}\bcLn{-1}{\bfg,\Lambda}\, \frZ_\Sigma\)
-\cR\Re\(\cLn{0}{\bfg,\Lambda}\de z^\Lambda\) \]
\nn\\
&&
-4\Im\frJ
\[\Re\( N^{\Lambda\Sigma}\bcLn{-1}{\bfg,\Lambda}\frZ_\Sigma\)
+\cR\Im\(\cLn{0}{\bfg,\Lambda}\de z^\Lambda\) \]
\nn\\
&&
+\frac{\cR}{2}\,\scK_\bfg\Biggl[\frac{\frS^2}{4(r+2c)^2}-\frac{(\de r)^2}{4(r+c)^2}
+\frac{\de r\, \de\log K}{2(r+c)}-\frac{\cA_K\frS}{r+2c}
-\frac{d K^2}{4K^2}+\cA_K^2\Biggr]
\nn\\
&&
-\biggl(\frac{\frS}{r+2c}-2\cA_K\biggr)
\biggl[N^{\Lambda\Sigma}
\Re\biggl(\Bigl(z^\Xi\scKn{1}{\Xi\bar\Sigma}+\bz^\Xi\scKn{-1}{\bar\Xi\bar\Sigma}\Bigr)\frZ_\Lambda\biggr)
-\cR\Im\Bigl(\Bigl(z^\Lambda\scKn{2}{\Lambda\Sigma}-\bz^\Lambda\scKn{0}{\bar\Lambda\Sigma}\Bigr)\de z^\Sigma\Bigr)\biggr]
\nn\\
&&
+\Bigl(\scKn{0}{\Lambda\bar\Sigma}+\scKn{0}{\bar\Sigma\Lambda}\Bigr)
\biggl(\frac{1}{\cR}\,N^{\Lambda\Xi} N^{\Sigma\Theta} \frZ_\Theta \bar\frZ_\Xi-\cR\de z^\Lambda \de\bz^\Sigma\biggr)
+2\I N^{\Lambda\Sigma}\Bigl(\scKn{1}{\Sigma\Xi}\,\bar\frZ_\Lambda\de z^\Xi-\scKn{-1}{\bar\Sigma\bar\Xi}\,\frZ_\Lambda \de\bz^\Xi\Bigr)
\Biggr\}
\nn\\
&-&
\frac{\I}{2}\,\de\log K\(z^\Lambda\frLn{1}{\Lambda}+\bz^\Lambda\bfrLn{-1}{\Lambda}\)
-\frac{2\I}{\cR}\(\frac{c\,\frS}{r+2c}+2r \cA_K\)\de\cIn{0}_\bfg
\nn\\
&+&
\frac{\I}{\cR}\(\de\cGn{0}{\Lambda}_\bfg\de\tzeta_\Lambda-\de\tcGn{0}{\bfg,\Lambda} \de \zeta^\Lambda\)
- \frac{\cR}{r}\, \Re\[z^\Lambda\frLn{0}{\Lambda}\,\bar\frJ\].
\label{frVtot1}
\eea
Note that the variable $\cR$ appears only in the instanton terms and therefore in our approximation it can be expressed
through $r$ using the perturbative relation \eqref{pert-r}.

The result for the D-instanton corrections given by the second term in \eqref{ds2final2}
can be compared with the linearization (in DT invariants $\bOm_\gamma$)
of the metric found in \cite{Alexandrov:2014sya}. We do not provide any details of this simple exercise which
shows a perfect match between the two metrics.

The result for NS5-instanton corrections given by the last term in \eqref{ds2final2} is new.
Note that the specific form of the function $\Psi_{\hbfg}(\xi)$ appearing in \eqref{5pqZ2} has not been used and
it is valid for any such function ensuring convergence of the corresponding integrals.
Unfortunately, the substitution of \eqref{Psibfg} does not seem to simplify the resulting metric
which still looks very complicated. But this is expected given that NS5-instantons break all continuous isometries
of the moduli space. It is also does not exhibit any fibration or other nice geometric structure.
Nevertheless, as will be shown in section \ref{sec-smallgs}, in the small string coupling limit, if one neglects terms
proportional to the differential of the instanton action, $\frV_\bfg$ reduces to the square of a one-form,
and this is precisely the structure expected from the analysis of string amplitudes.

\section{Universal hypermultiplet}
\label{sec-UHM}

The case of a rigid CY, i.e. without complex structure moduli ($h^{2,1}(\CY)=0$), is special.
Then the spectrum contains only the {\it universal hypermultiplet} \cite{Strominger:1997eb}
comprising the dilaton $r$, NS axion $\sigma$  and a pair of RR fields\footnote{We drop
indices on quantities labelled by $\Lambda,\Sigma,\dots$ since in this case that take a single value.}
$\zeta,\tzeta$, so that the moduli space is four-dimensional. Such QK manifolds are known to
have an alternative description due to Przanowski \cite{Przanowski:1984qq}
in terms of solutions of a certain non-linear partial differential equation.
This fact allows to test our one-instanton corrected metric which must be compatible with this description.
In particular, it must produce a solution of the linearized differential equation.
Furthermore, relating this solution to known solutions in the literature, one may hope to find an alternative
twistorial formulation of NS5-instantons that is better adapted to symmetries of type IIA theory.
In this section we show that the metric \eqref{ds2final2} is indeed consistent with the Przanowski description,
obtain the corresponding potential solving the differential equation, but leave the problem of relating it to
other solutions for future work.

\subsection{Przanowski description}
\label{subsec-Prz}

In \cite{Przanowski:1984qq} it was proven that locally, on a four-dimensional self-dual Einstein space
(which is a characterization of QK geometry in four dimensions) with a negative curvature,
one can always find complex coordinates $z^\alpha$ ($\alpha=1,2$) such that the metric takes the following form
\be
\de s^2_z[h] = -\frac{6}{\Lambda}\(h_{\alpha \bar \beta}\de z^\alpha \de \bz^\beta + 2 e^h |\de z^2|^2\),
\label{metPrz}
\ee
where $h_{\alpha}= {\partial h}/{\partial z^\alpha}$, etc. This metric is completely determined by
a single real function $h(z^\alpha, \bz^\alpha)$ that must satisfy the following non-linear partial differential equation
\bea
\Prz_z[h]\equiv h_{1\bar1}h_{2\bar2} - h_{1\bar2}h_{\bar1 2}+\(2h_{1\bar1}-h_{1}h_{\bar1}\)e^{h}=0.
\label{Prz-master-equation}
\eea

Of course, the equation is too complicated to be solved in general. However, the problem significantly simplifies
if one already knows a solution $\hpert$ describing some QK space and one is interested in linear deformations of this space.
The point is that such deformations are governed by the {\it linearization} of \eqref{Prz-master-equation}
around $\hpert$, and a linear equation is much easier to solve.
This is precisely the case for $\cM_H$ whose perturbative metric has a well known Przanowski description \cite{Alexandrov:2006hx}
in terms of coordinates
\be
z^1=-\bigl(r+c\log(r+c)\bigr)-\frac{\I}{4}\, (\sigma+\zeta\tzeta+\tau\zeta^2),
\qquad
z^2=\frac{\I}{2}\, (\tzeta+\tau\zeta)
\label{pert-zz}
\ee
and the Przanowski potential
\be
\hpert=-\log\frac{2\tau_2 r^2}{r+c}\, ,
\label{pert-h0}
\ee
where $\tau\equiv \tau_1+\I\tau_2$
is a fixed complex parameter with $\tau_2>0$, given by the ratio of periods of the holomorphic 3-form $\Omega\in H^{3,0}(\CY)$
over an integral symplectic basis $(A,B)$ of $H_3(\CY,\IZ)$, $\tau=-{\int_B\Omega}/{\int_A\Omega}$,
which defines the holomorphic prepotential in the rigid case \cite{Bao:2009fg}
\be
F(X) = -\frac{\tau}{2}\, X^2.
\label{prepUHM}
\ee
It is easy to check that with these definitions the equation \eqref{Prz-master-equation} is satisfied and
the metric \eqref{metPrz} reproduces \eqref{1lmetric} provided one sets the cosmological constant to be $\Lambda=-3/2$.
The linearization of the Przanowski equation around this background takes the form  \cite{Alexandrov:2006hx,Alexandrov:2009vj}
\be
\bigl(\Delta+1\bigr)\( r^2\delta h\)=0,
\label{linPrz}
\ee
where $\Delta$ is the Laplace-Beltrami differential operator defined by the perturbative metric
and we expanded $h=\hpert+\delta h$ keeping only terms linear in $\delta h$.
An explicit expression of $\Delta$ in terms of the real coordinates is recorded in \eqref{Laplacereal}.
Solutions of \eqref{linPrz} have been studied in \cite{Alexandrov:2006hx}.
We discuss them below in section \ref{subsec-relPrz} in relation to our results.
But before, we demonstrate that our metric does fit the Przanowski description, which can be considered as a non-trivial test
on our calculations.

\subsection{The metric and Przanowski potential}

In \cite{Alexandrov:2009vj} it was shown that the Przanowski description follows directly from the twistorial construction
for a generic four-dimensional QK space.
Although this description is not unique because there is a large ambiguity in the choice of coordinates $z^\alpha$
(which also affects the Przanowski potential $h$),
it was found that a particularly convenient choice is given by
\be
z^1 = \frac{\I}{2}\,\ai{+}_0 -2c \log \xi_{-1},
\qquad
z^2 = \frac{\I}{2}\,\txii{+}_0,
\label{genz}
\ee
where in our case $\xi_{-1}=\cR$, $\ai{+}$ and $\txii{+}$ are defined in \eqref{redefDC},
and their Fourier coefficients are evaluated in \eqref{LaurentDarboux}.
The Przanowski potential should then be equal to
\be
h=-2\phi+2\log(\xi_{-1}/2)= -2\log \frac{2r}{\cR}\, .
\label{genh}
\ee
Thus, it is sufficient to plug in these identifications into \eqref{metPrz} and to verify that
the resulting metric reproduces \eqref{ds2final2}. Similarly, one can check that the Przanowski equation \eqref{Prz-master-equation}
is also satisfied. Although we have performed these checks, the corresponding calculations are extremely cumbersome
because one has to evaluate derivatives of instanton corrected functions with respect to instanton corrected coordinates.
Therefore, we prefer to present an alternative derivation which, on one hand, avoids this complication and, on the other hand,
establishes a contact with solutions of the linearized Przanowski equation discussed in the previous subsection.

The idea is that at the linearized level the variation of the metric should be encoded in the variation
of the Przanowski potential $\delta h(z)$.
From \eqref{metPrz} one finds that it induces the following linear deformation of the metric
\be
\delta_h \de s^2=4\delta h_{\alpha\bar\beta} \de z^\alpha\de\bz^\beta+8 e^{\hpert} \delta h |\de z^2|^2.
\label{linPrzmet}
\ee
However, this assumes that the original non-deformed metric is written in terms of the non-deformed complex coordinates $z^\alpha$.
However, in practice these coordinates are also deformed as functions of a fixed set of real coordinates $\vph^m$ which in our case
coincide with $(r,\sigma,\zeta,\tzeta)$.
And it is these real coordinates that are used to define the non-deformed metric.

In such situation, the variation of the metric gets an additional contribution. To find it, let $(z^\alpha(\vph),h(z))$
and $(\zpert^\alpha(\vph),\hpert(\zpert))$ denote the deformed and non-deformed complex coordinates
and Przanowski potential. In our case, they are given in \eqref{genz}, \eqref{genh} and \eqref{pert-zz}, \eqref{pert-h0}, respectively.
By construction, we have $\Prz_z[h]=\Prz_{\zpert}[\hpert]=0$.
Using these functions, we further define
\be
\delta_\vph h=h(z(\vph))-\hpert(\zpert(\vph)).
\ee
Note that this variation of the Przanowski potential is different from the one defined above
which reads $\delta h(z)=h(z)-\hpert(z)$ and satisfies the linearized Przanowski equation.
The relation between these two functions is obtained as follows
\be
\delta h=\delta_\vph h-\(\hpert(z(\vph))-\hpert(\zpert(\vph))\)
=\delta_\vph h -\hpert_\alpha \delta z^\alpha- \hpert_{\bar\alpha} \delta \bz^\alpha,
\label{res-deltah}
\ee
where we introduced $\delta z^\alpha(\vph)=z^\alpha(\vph)-\zpert^\alpha(\vph)$.

Now we can write the full deformation of the metric which keeps the coordinates $\vph^m$ fixed as
\be
\delta \de s^2=\de s^2_{z(\vph)}[h]-\de s^2_{\zpert(\vph)}[\hpert]
=\delta_h \de s^2+\delta_z\de s^2,
\label{deltamet}
\ee
where the first term is defined in \eqref{linPrzmet} and
\be
\delta_z \de s^2=\de s^2_{z(\vph)}[\hpert]- \de s^2_{\zpert(\vph)}[\hpert]
\label{var-metric}
\ee
is the deformation of the Przanowski metric defined by the {\it non-deformed} potential $\hpert$
under the variation of the complex coordinates as functions  of the fields $\vph^m$.
Using \eqref{metPrz} with $\Lambda=-3/2$ and taking into account that our non-perturbed potential \eqref{pert-h0} satisfies
$\hpert_\alpha=\hpert_{\bar \alpha}$,
$\hpert_{\alpha \bar \beta}=\hpert_{\alpha\bar\beta}$, etc., we find an explicit formula
\be
\begin{split}
\delta_z \de s^2
=&\, 8\hpert_{\alpha \beta\gamma}\Re(\delta z^\gamma)\de \zpert^\alpha \de \bzpert^\beta
+8\hpert_{\alpha \beta}\Re(\de\delta z^\alpha \de\bzpert^\beta)
\\
&\, + 16 e^{\hpert}\( \hpert_\alpha \Re(\delta z^\alpha)|\de \zpert^2|^2
+\Re(\de\delta z^2 \de \bzpert^2)\).
\end{split}
\label{coord-Przmt}
\ee

To apply these results to the case under consideration, we obtain from \eqref{genz} and \eqref{genh}
the variation of the complex coordinates
\bea
\delta z^1 &=&
- \frac{1}{32\pi^3}\sum_{\gamma}\sigma_\gamma \bOm_\gamma \(\cJn{0}_\gamma+\frac{2\pi \I \cR r}{r+c}\, Z_\gamma \cJn{1}_\gamma\)
\nn\\
&&
- \frac{1}{32\pi^3}\sum_\bfg \bOm_\gamma\(\(1-\frac{4\pi k c r}{r+c}\) \cIn{0}_\bfg
-\frac{2\pi \I k\cR r}{r+c}\,\scLn{1}_\bfg \)
-\zeta\delta z^2,
\label{Prz-delta-cplxcoord}\\
\delta z^2 &=& \frac{\I}{16\pi^2}\sum_\gamma \sigma_\gamma \bOm_\gamma Z_\gamma\cJn{0}_\gamma
- \frac{\I}{16\pi^2}\sum_\bfg \bOm_\gamma k \,\scLn{0}_\bfg,
\nn
\eea
and the Przanowski potential
\bea
\delta_\vph h&=&\delta_\vph \log\cR^2
\\
&=&
-\frac{\I\cR}{32\pi^2(r+c)}\[\sum_\gamma \sigma_\gamma\bOm_\gamma \( Z_\gamma \cJn{1}_\gamma+\bZ_\gamma\cJn{-1}_\gamma\)
-\sum_\bfg \bOm_\gamma\, k\(\scLn{1}_\bfg+\bscLn{-1}_\bfg\)\],
\nn
\eea
where we used \eqref{fullr} to get $\delta_\vph \cR$.
Then due to \eqref{res-deltah} and \eqref{firstderh}, we have
\be
\delta h=  \delta_\vph h -\frac{1}{r}\Re\(\delta z^1+\zeta \delta z^2\)
= \frac{1}{32\pi^3 r}\[\sum_{\gamma}\sigma_\gamma \bOm_\gamma \,\cJn{0}_\gamma
+\sum_\bfg \bOm_\gamma\, \cIn{0}_\bfg\].
\label{ourdeltah}
\ee
This result is perfectly consistent with the general formula obtained in \cite{Alexandrov:2009vj} for the solution
of the linearized Przanowski equation \eqref{linPrz} corresponding to the deformation induced by a set of
transition functions $H_i$
\be
\delta h=\frac{1}{8\pi r} \sum_i \int_{C_i} \frac{\de t}{t}\, H_i.
\label{delh}
\ee
The fact that it satisfies \eqref{linPrz} follows from the identity \cite[Eq.(6.15)]{Alexandrov:2009vj}
\be
\[\Delta+1+\frac{r^2}{r+c}\,\p_t\(\frac{t}{4(r+2c)}\,(t\p_t+1)+4\I t\p_\sigma \)\]
\(\frac{r}{t}\, H\)=0
\ee
valid for any holomorphic function $H$ of the perturbative Darboux coordinates \eqref{exline},
which reduces the action of the Laplace-Beltrami operator to a total derivative.

The last step is to check that the deformation of the metric \eqref{deltamet} coincides with
the last two terms in \eqref{ds2final2}. Since all coefficients in \eqref{linPrzmet} and \eqref{coord-Przmt}
are given by derivatives of the non-perturbed potential $\hpert$, the computation is straightforward
and confirms the agreement. We present some details of this computation in appendix \ref{ap-UHM}.

\subsection{Relation between different solutions}
\label{subsec-relPrz}

As was mentioned in section \ref{subsec-Prz}, solutions of the linearized Przanowski equation \eqref{linPrz}
were investigated in \cite{Alexandrov:2006hx}. As a result, it was found that physically relevant solutions,
i.e. having exponential dependence on the inverse string coupling characteristic for D-brane or NS5-brane instantons,
are generated by two sets of functions\footnote{In \cite{Alexandrov:2006hx} the parameter $\tau$ was fixed to be $\tau=\I/2$,
but it is easy to generalize the results to its arbitrary value.
Besides, the analysis of \cite{Alexandrov:2006hx} also produced a third solution
which however does not have an obvious interpretation. Due to this, we omit it here.}
\bea
\delta h_{p,q}& =& \frac{1}{r}\,e^{-2\pi \I(q\zeta-p\tzeta)} K_0\left( 4\pi|q+\tau p|\sqrt{{2(r+c)}/{\tau_2}} \right) ,
\label{d2}
\\
\delta h_{k}&=& e^{-\pi \I k \sigma-\frac{\pi |k|}{2\tau_2}\,|\tzeta+\tau\zeta|^2} \, Z_{|k|}(r),
\label{solns5}
\eea
where $K_0(x)$ is the modified Bessel function and
\be
Z_k(r)=\frac{e^{4\pi k r}}{r (r+c)^{4c\pi k}}\,
\int_1^{\infty}e^{-8\pi k(r+c)t}\frac{{\rm d}t}{t^{1+8\pi ck}}
\label{defZr}
\ee
can be expressed through the incomplete Gamma function. A general solution can be obtained as a linear combination of the
Heisenberg transformations \eqref{Heis} applied to the basis solutions \eqref{d2} and \eqref{solns5}.
Since these transformations do not affect the functional form of the first solution, we can write
\be
\delta h=
\sum_{p,q} C_{p,q}  \delta h_{p,q}+\sum_{k\ne 0}
\int \de \eta^\Lambda \int\de\tleta_\Lambda  \, C_{k}(\eta,\tleta)\, T_{\eta^\Lambda,\tleta_\Lambda,0}\[\delta h_{k}\].
\label{genlinsol}
\ee
It is clear that the first term represents D-instanton corrections of charge $\gamma=(p,q)$, while the second term
is the effect of NS5-instantons.

Which constants $C_{p,q}$ and functions $ C_{k}(\eta,\tleta)$
correspond to the physical metric on $\cM_H$ cannot be determined from the Przanowski framework and requires an additional input.
Such input is provided by our result \eqref{ourdeltah} for the linear deformation of the Przanowski potential.
Comparing it with \eqref{genlinsol}, it should be possible to find the coefficients $C_{p,q}$ and the functions $ C_{k,\eps}(\eta,\tleta)$.
For example, matching the first terms in these expressions with help of \eqref{cJnK}, one obtains
\be
C_{p,q}=\frac{\sigma_\gamma\bOm_\gamma}{16\pi^3}\, .
\ee

However, a similar match for the solutions representing NS5-instantons seems to be much more complicated.
It requires finding $ C_{k}(\eta,\tleta)$ such that
\be
e^{-\pi \I k \sigma}\,Z_{|k|}(r)\int \de \eta^\Lambda \int\de\tleta_\Lambda  \, C_{k}(\eta,\tleta)\,
e^{-\frac{\pi |k|}{2\tau_2}\left|\tzeta+\tleta+\tau(\zeta+\eta)\right|^2+\pi\I k (\zeta\tleta-\tzeta\eta)}
=\frac{1}{32\pi^3 r}\sum_{p,\hgam} \bOm_\gamma\, \cIn{0}_\bfg.
\label{matchNS5}
\ee
First, it is not clear why the dependence on $r$ should factorize. If this does not happen on the r.h.s.,
the match appears to be impossible. Second, while the integral \eqref{defZr}
has some similarities to the twistorial integrals representing instanton corrections, its integration contour is quite different
and it is not clear how it can emerge in our context.
Thus, either a kind of Poisson resummation over a subset of charges on the r.h.s. of \eqref{matchNS5}
produces these features or (which is probably more likely) the analysis of \cite{Alexandrov:2006hx}
missed more general solutions to the linearized Przanowski equation by imposing too strong conditions selecting
physically relevant solutions. We leave a resolution of this puzzle to future work.

\section{Small string coupling limit}
\label{sec-smallgs}

In this section we extract the small string coupling limit of the one-instanton corrections to the hypermultiplet metric
that we calculated in section \ref{sec-metric}.
This is the limit where we expect to establish a connection with string amplitudes.
For D-instantons this has already been done in \cite{Alexandrov:2021shf},
therefore here we concentrate on NS5-instantons.
First, we obtain a general structure of the instanton corrected metric following from analysis
of the effective action and string amplitudes. Then we show that exactly the same structure emerges
in the small string coupling limit of the metric \eqref{ds2final2}, thereby providing predictions
for a certain class of string amplitudes in NS5-brane background.
This limit however leads to a somewhat unusual instanton action.
So in the last subsection we consider an additional limit of small RR fields which allows us to recover the standard action
and crucially simplifies our predictions for the amplitudes.

\subsection{Instantons from string amplitudes}

The analysis of this subsection is very similar to the one in \cite[\$5 and \$6.1]{Alexandrov:2021shf}
so we will be more brief.

Our goal is to relate the metric on the hypermultiplet module space to scattering amplitudes of physical fields.
Since the relevant fields are massless scalars, the first non-trivial amplitudes are 4-point functions.
Therefore, we need to generate a 4-point interaction vertex from a metric dependent term in the effective action.
The simplest possibility is to consider the kinetic term for hypermultiplet scalars $\vph^m$ parametrizing $\cM_H$
\be
-{1\over 2} \, \int \de^4 x\( g^{\rm pert}_{mn} +  \sum_\bfg \, e^{-S_\bfg}\( h^{(\bfg)}_{mn}(\vph)+\cdots \)\)
\p_\mu \vph^m \p^\mu \vph^n .
\label{escalar}
\ee
Here we substituted the expected form of the metric in the small string coupling limit where
it is equal to the perturbative metric plus instanton corrections proportional to the exponential of the instanton action.
We kept only NS5-instanton contributions, denoted NS5-instanton action by $S_\bfg$ and
the leading term in the expansion of the tensor multiplying the exponential by $h^{(\bfg)}_{mn}$.
In the limit $g_s\to 0$, we expect that $S_\bfg\sim g_s^{-2}$ and
assume that the fields $\vph^m$ are normalized so that they stay constant.

Let us now expand the fields around their expectation values $\phi^m$. If $\lambda^m=\vph^m-\phi^m$ denotes the fluctuations,
then the expansion of \eqref{escalar} generates infinitely many interaction vertices for these fluctuations. In particular,
the leading instanton contribution to the $\lambda^4$-term is obtained by bringing down
two factors of $\lambda^m$ from the expansion of the instanton action and is given by
\be
-{1\over 4} \sum_\bfg \int \de^4 x\,  e^{-S_\bfg(\phi)}\, \p_p S_\bfg(\phi) \, \p_q S_\bfg(\phi) \,
 h^{(\bfg)}_{mn}(\phi)\,  \lambda^p \,
\lambda^q \, \p_\mu\lambda^m \p^\mu\lambda^n .
\label{eleadscalar}
\ee
This term induces an instanton contribution to 4-point functions of fields $\lambda^{m_i}$ which reads as
\be
 (2\pi)^4 \delta^{(4)}\left(\sum_i p_{i}\right) e^{-S_\bfg}
\bigg[\p_{m_1} S_\bfg\,  \p_{m_2} S_\bfg \,  h^{(\bfg)}_{m_3 m_4}\,  p_{34}
+\ {\mbox{inequivalent perm.} \atop \mbox{of 1,2,3,4}}
\bigg]\, ,
\label{eampin}
\ee
where $ p_{i}^\mu$ is the momentum carried by $\lambda^{m_i}$ and $p_{ij}=\eta_{\mu\nu} p_{i}^\mu p_{j}^{\nu}$.

The amplitude \eqref{eampin} induced by the term \eqref{eleadscalar} in the effective action is to be compared with
the explicit computation of the instanton amplitude in string theory.
First, we note that NS5-instanton perturbation theory is similar to the one for D-instantons
\cite{Sen:2020cef}, but with open string diagrams ending on D-branes replaced by closed string diagrams
in the presence of an NS5-brane. In particular, the instanton action should be given by the sphere diagram
in the NS5-background which agrees with its scaling as $g_s^{-2}$.
The overall normalization factor should be given by the exponential of the torus diagram,
and each insertion of a closed string vertex operator,
at leading order in $g_s$, gives rise to a factor given by the sphere one-point function of this operator.

This is not the end of the story, however, since the instanton breaks half of the $N=2$ supersymmetry.
The four broken supercharges imply the existence of four Goldstino zero modes.
To get a non-vanishing result from integration over these modes, their vertex operators should
be inserted in the sphere diagrams composing our amplitude.
As a result, schematically, the NS5-instanton contribution to the 4-point function we are interested in is given by
\be
(2\pi)^4 \delta^{(4)}\left(\sum_i p_{i}\right)\bOm_\gamma \cN_\bfg\, e^{-S_\bfg}
\int \[\prod_{\alpha,\dot\alpha=1,2}\de\chi^\alpha\de\chi^{\dot\alpha}\] A_{\bfg,\alpha\dot\alpha\beta\dot\beta}^{m_1m_2m_3m_4}
\chi^\alpha\chi^{\dot\alpha}\chi^\beta\chi^{\dot\beta},
\label{gen4point}
\ee
where $\cN_\bfg$ is the normalization factor computed by torus with removed zero modes,
$\chi^\alpha$ are the fermionic zero modes, and  $A_{\bfg,\alpha\dot\alpha\beta\dot\beta}^{m_1m_2m_3m_4}$ is a sum of products
of four sphere diagrams, each with one closed string vertex operator corresponding to one of $\lambda^{m_i}$,
and four fermion zero modes distributed among the four spheres.
We also included the factor of $\bOm_\gamma$ which for primitive $\gamma$ counts the number of BPS instantons in a given homology class,
and for non-primitive charges takes also into account multi-covering effects. The fact that these effects combine to give the rational
DT invariant \eqref{defntilde} can be argued in the same way as for D-instantons \cite{Alexandrov:2021shf}.

This expression can be further simplified, because each sphere diagram must carry even number of fermion zero modes.
Hence only two situations are possible: either all four zero modes are inserted on one sphere, or
two spheres carry two zero modes each and two spheres are without them.
Moreover, one can argue that the former configuration does not contribute.
Indeed, the sphere one-point function of the vertex operator corresponding to $\lambda^m$,
without additional insertions of the fermion zero modes, is simply given by
the derivative of the instanton action $-\p_m S_\bfg$.
In particular, it does not depend on the momentum carried by the vertex operator.
Therefore, all momentum dependence in the case where all four fermion zero modes are
inserted on a single sphere comes from this sphere diagram.
However, the Lorentz invariance implies that it should be a function of $p^2$
where $p$ is the momentum carried by the vertex operator on this sphere. But since
$p^2=0$, this contribution does not depend on $p_{i}$'s at all and would give rise
to a potential term in the effective action.
Since instanton corrections should not generate any potential, this amplitude is expected to vanish.

Thus, the only surviving contribution is the one where we have two of the zero modes on
one sphere, two on another sphere, and two spheres without zero modes which, as we already noted, produce
the factors $-\p_m S_\bfg$. Let us estimate the sphere diagram with the zero mode insertions.
Note that to have a non-vanishing coupling with the momentum vector, one of the zero modes must carry dotted index
and the other one should carry undotted index. Then the full diagram can be represented as
\be
\I \cA^{(\bfg)}_m(\phi) \, p_\mu\, \gamma^\mu_{\alpha\dot\alpha}  ,
\label{eterm}
\ee
where $\cA^{(\bfg)}_m(\phi)$ is a function of background fields independent of the momentum.
Collecting all contributions, we find that
\be
A_{\bfg,\alpha\dot\alpha\beta\dot\beta}^{m_1m_2m_3m_4}=
-\p_{m_1} S_\bfg\,  \p_{m_2} S_\bfg \,
\Bigl(\cA^{(\bfg)}_{m_3}\, p_{3,\mu}\, \gamma^\mu_{\alpha\dot\alpha}\Bigr)
\Bigl( \cA^{(\bfg)}_{m_4} \, p_{4,\nu}\, \gamma^\nu_{\beta\dot\beta} \Bigr)
+\ {\mbox{inequivalent perm.} \atop \mbox{of 1,2,3,4.}}
\ee
Substituting this result into \eqref{gen4point}, integrating over the zero modes, and using that
$\eps^{\alpha\beta}\eps^{\dot\alpha\dot\beta}\,\gamma^\mu_{\alpha\dot\alpha} \,\gamma^\nu_{\beta\dot\beta}
=-\Tr(\gamma^\mu\gamma^\nu)=-2\, \eta^{\mu\nu}$,
one obtains that the NS5-instanton contribution to the 4-point function has
the following form\footnote{We are sloppy here about numerical factors.
Moreover, as shown in \cite{Alexandrov:2021shf}, there is also an additional factor that must be taken into account coming from
a difference between the four-dimensional metric in the string frame used to calculate string amplitudes
and in the frame used to write the effective action \eqref{escalar} with vector and hypermultiplets decoupled.
We assume that all such factors have been absorbed into $\cN_\bfg$.}
\be
(2\pi)^4 \delta^{(4)}\left(\sum_i p_{i}\right)\bOm_\gamma \cN_\bfg\, e^{-S_\bfg}
\bigg[\p_{m_1} S_\bfg\,  \p_{m_2} S_\bfg \,  \cA^{(\bfg)}_{m_3}\cA^{(\bfg)}_{m_4}\,  p_{34}
+\ {\mbox{inequivalent perm.} \atop \mbox{of 1,2,3,4}}
\bigg].
\label{gen4point-final}
\ee
Comparing \eqref{gen4point-final} with \eqref{eampin}, one finds that they have exactly the same structure.
This allows to extract the metric $h^{(\bfg)}_{mn}(\vph)$:
\be
h^{(\bfg)}_{mn}=\bOm_\gamma \cN_\bfg \cA^{(\bfg)}_m\cA^{(\bfg)}_n.
\label{e56}
\ee

However, the above argument is not quite exact because it is insensitive to the terms in the action \eqref{escalar}
proportional to $\p_m S_\bfg$. Indeed, such terms can be generated either by integration by parts or by
a change of variables involving non-perturbative terms \cite{Alexandrov:2021shf}.
In either case the scattering amplitudes should not be affected and hence \eqref{e56} is valid only up to addition
of the gradient of the instanton action.

To recapitulate, it is convenient to use the language of differential forms.
Let us define $\cA_\bfg=\cA^{(\bfg)}_m(\vph)\de\vph^m$. Then the above analysis of string amplitudes shows that
in the small $g_s$ limit the NS5-instanton contribution to the hypermultiplet metric should be of the form
\be
\de s^2_{\rm NS5}\simeq
\sum_\bfg \bOm_\gamma \cN_\bfg\, e^{-S_\bfg}\( \cA_\bfg^2+\cB_\bfg \de S_\bfg\)
\label{predictNS5}
\ee
with some one-form $\cB_\bfg$ which this analysis cannot fix.
Below we verify that the metric \eqref{ds2final2} does fit this form and find all functions and one-forms
appearing in \eqref{predictNS5} explicitly. On one hand, this provides another non-trivial check on our results,
and on the other hand, gives a prediction for the amplitudes $\cA^{(\bfg)}_m$.

\subsection{The limiting metric}

\subsubsection{Definition of the limit}

Before extracting the small string coupling limit, we should properly definite it.
Namely, we should specify how various fields behave in this limit. Naively, it is enough to send
the variable $r$, related to the dilaton, to infinity and to keep all other variables fixed.
In particular, it is this naive limit that has been used in \cite{Alexandrov:2010ca}
to extract the NS5-brane instanton action\footnote{The analysis of \cite{Alexandrov:2010ca} ignored the effect of
the logarithmic term in $\talp$ \eqref{exline} incorporating the one-loop $g_s$-correction.
In appendix \ref{ap-Sinst} we show that its effect on the saddle point evaluation of a typical integral
describing NS5-instanton effects is quite non-trivial, but affects only the fluctuation determinant around
the instanton leaving the instanton action \eqref{NS5instact} intact.}
that, on one hand, reproduces the result of a classical analysis of instanton solutions in $N=2$ supergravity \cite{deVroome:2006xu}
and, on the other hand, makes contact with the Gaussian NS5-partition function obtained by holomorphic factorization
\cite{Witten:1997hc,Belov:2006jd}:
\be
S_\bfg^{(0)}=4\pi k r +\pi\I k \left(  \sigma + \zeta^\Lambda \tzeta_\Lambda
-2 n^\Lambda \tzeta_\Lambda
- \bar\cN_{\Lambda\Sigma} (\zeta^\Lambda-n^\Lambda) (\zeta^\Sigma-n^\Sigma)\right)
-2\pi \I m_\Lambda z^\Lambda.
\label{NS5instact}
\ee
Here
\be
\cN_{\Lambda \Sigma} = \bF_{\Lambda \Sigma}
-\I\, \frac{ N_{\Lambda \Lambda'}z^{\Lambda'} N_{\Sigma \Sigma'}z^{\Sigma'}}{z^\Theta N_{\Theta \Theta'}z^{\Theta'}}
\ee
is the matrix of the gauge couplings in the mirror type IIB formulation, and the formula \eqref{NS5instact} is valid
for positive $k$, while for negative $k$ the first term flips the sign
and $\bar\cN_{\Lambda\Sigma}$ should be replaced by $\cN_{\Lambda\Sigma}$.
This ensures the convergence of the sum over $n^\Lambda$ for both signs of the NS5-brane charge
due to the physical condition $\Im \cN_{\Lambda\Sigma}<0$.

However, this naive limit suffers from a problem.
It is easy to see already for the classical metric obtained from \eqref{1lmetric} by setting $c=0$
that different terms have different scaling in $r$.
This makes it difficult even to formulate what is meant by the leading order metric in the
large $r$ limit.
On the other hand, in \cite{Alexandrov:2021shf} it was noticed that one does get a homogeneous scaling in $g_s$
for both the classical metric and the small string coupling limit of D-instanton corrections provided we take this limit
as
\be
r,\sigma \sim g_s^{-2},
\qquad
\zeta^\Lambda,\tzeta_\Lambda \sim g_s^{-1},
\qquad
z^a \sim g_s^0,
\qquad
g_s\to 0,
\label{scaling}
\ee
which also implies $\cR\sim g_s^{-1}$.
Besides, in this modified limit the D-instanton corrections have been shown to acquire essentially
the same quadratic structure as in \eqref{predictNS5}
and matched exactly against computations of string amplitudes.
This strongly suggests that \eqref{scaling} is the correct limit to consider for NS5-instantons as well.

In fact, the origin of the scaling \eqref{scaling} can be easily understood from the supergravity action in
ten dimensions\footnote{We thank Ashoke Sen for clarification of this issue.}
where the kinetic terms in the NS sector are multiplied by the factor $e^{-2\phi_{(10)}}\sim g_s^{-2}$. Since such a factor is absent in
the RR kinetic terms, the RR fields should scale as $g_s^{-1}$ so that the whole action scales uniformly.
Finally, the scaling of $\sigma$ follows from the dualization of the $B$-field.
Moreover, this rescaling of the RR fields is necessary to match them with their worldsheet counterparts \cite{Polchinski:1998rr}.
Therefore, it is also necessary to establish a correspondence between the small $g_s$ expansion of the effective action
and the genus expansion of string theory. Hence, if we want to derive predictions for any string amplitudes,
we must study the limit \eqref{scaling} rather than the naive one where only $r$ scales with $g_s$.
Below we do it for the NS5-instanton contribution to the metric \eqref{ds2final2}.

\subsubsection{Saddle point evaluation}

As an important preliminary step, let us evaluate in the small $g_s$ limit, as it is defined in \eqref{scaling},
the following integral
\be
\int_{\ell_\bfg}\frac{\de t}{t}\, f(t) \, e^{-2\pi\I k\cS_\bfg(t)},
\label{intft}
\ee
where $f(t)$ is a polynomial in $t$ and $t^{-1}$,
\be
\cS_\bfg(t)=\frac{1}{2} \(\talp+(\xi^\Lambda-2n^\Lambda) \txi_\Lambda\)
- F(\xi-n)-\frac{m_a(\xi^a-n^a) +Q}{k(\xi^0-n^0)}-\frac{m_0}{k}\, ,
\label{effS}
\ee
and all Darboux coordinates in \eqref{effS} are set to their perturbative expressions \eqref{exline}.
This type of integrals multiplies all terms in \eqref{frVtot1} with positive $k$
and thus encodes NS5-instanton corrections to the hypermultiplet metric.

In the limit the ``effective action" $\cS_\bfg (t)$ can be expanded as $\cS_\bfg(t)=-4\I c  \log t +\sum_{\ell\ge 0} \cS_{\bfg,\ell}$
where $\cS_{\bfg,\ell}$ scales as $g_s^{\ell-2}$ and we extracted the only term having a logarithmic dependence on $t$.
Note that the expansion starts from the term scaling as $g_s^{-2}$,
as is expected for NS5-instantons.
For our purpose, it is sufficient to keep in the exponential only terms with non-positive scaling power, i.e. with $\ell=0,1,2$.
Then the resulting integral can be evaluated by saddle point.
It is easy to see that at the leading order the result is given by
\be
\frac{f(t_0)\, e^{- S_\bfg}}{t_0^{1+8\pi k c}\sqrt{\I k\cS''_{\bfg,0}}}\, ,
\qquad
S_\bfg=2\pi\I k\(\cS_{\bfg,0}+ \cS_{\bfg,1}+\cS_{\bfg,2}-\hf\, \frac{(\cS'_{\bfg,1})^2}{\cS''_{\bfg,0}}\),
\label{saddleresult}
\ee
where all $\cS_{\bfg,\ell}$ and their derivatives (denoted by primes) are evaluated at $t_0$ which is a solution of
the leading order saddle point equation $\cS'_{\bfg,0}=0$.

From \eqref{effS}, we find that
\begin{subequations}
\bea
\cS_{0}(t)&=& \hf\(\sigma+\zeta^\Lambda\tzeta_\Lambda\)-\cR^2\Re(\bz^\Lambda F_\Lambda(z))
+ \cR\zeta^\Lambda \(t^{-1} F_\Lambda(z) -t \bF_\Lambda(\bz)\)
\nn\\
&&\qquad
+\cR^2\(t^{-2} F(z)+t^2 \bF(\bz)\)-F(\xi(t)),
\label{cS0}
\\
\cS_{\bfg,1}(t)&=& - n^\Lambda \[ \tzeta_\Lambda + \cR\(t^{-1}F_\Lambda(z) - t \bF_\Lambda(\bz)\) - F_{\Lambda}(\xi(t))\],
\label{Sg1}
\\
\cS_{\bfg,2}(t)&=&-\frac{1}{2}\,n^\Lambda n^\Sigma F_{\Lambda \Sigma}(\xi(t))-\frac{m_a}{k}\, \frac{\xi^a(t)}{\xi^0(t)}-\frac{m_0}{k}\, .
\eea
\end{subequations}
Note that we dropped the index $\bfg$ on $\cS_0$ because this part of the effective action does not depend on any charges.
Taking the first derivative of \eqref{cS0}, one finds that the equation on $t_0$ can be written as
\be
t_0^{-1} z^\Lambda \euF_\Lambda+t_0 \bz^\Lambda\beuF_\Lambda=-\I\cR K,
\label{eqt0}
\ee
where we introduced
\be
\begin{split}
\euF_\Lambda &= F_\Lambda(\xi(t_0)) - \xi^\Sigma(t_0) F_{\Lambda\Sigma}(z),
\\
\beuF_\Lambda &= F_{\Lambda}(\xi(t_0)) - \xi^\Sigma(t_0) \bF_{\Lambda\Sigma}(\bz).
\end{split}
\label{reverseSaddle}
\ee
Note that for generic prepotential this equation is highly non-linear and cannot be solved explicitly, while in the rigid case
($h^{2,1}(\CY)=0$) where $F(X)$ is quadratic and given by \eqref{prepUHM} one finds
$\euF_0=0$, $\beuF_0=-2\I\tau_2 \xi(t_0)$ and $t_0=\zeta/\cR$.
The second derivative appearing in \eqref{saddleresult} is found to be
\be
\begin{split}
t_0^2 \cS''_{0}(t_0)
=&\,
\cR\Bigl[ \zeta^\Lambda\( t_0^{-1} F_{\Lambda}(z)-t_0 \bF_\Lambda(\bz)\)  +4\cR \(t_0^{-2} F(z) +t_0^2 \bF(\bz)\)
\\
&\, -\(\zeta^\Lambda\(t_0^{-1}z^\Sigma-t_0\bz^\Sigma\) +2\cR\(t_0^{-2} z^\Lambda z^\Sigma+t_0^2 \bz^\Lambda\bz^\Sigma\)\)
F_{\Lambda\Sigma}(\xi(t_0))\Bigr].
\end{split}
\ee
Finally, the instanton action defined in \eqref{saddleresult} is given by
\bea
S_\bfg &=&
2\pi\I k\Biggl[
\hf\(\sigma+ \zeta^\Lambda \tzeta_\Lambda\) - \cR^2 \Re (\bz^\Lambda F_\Lambda)
+\cR \zeta^\Lambda (t_0^{-1} F_\Lambda-t_0 \bF_\Lambda)  + \cR^2 \(t_0^{-2} F +  t_0^2 \bF\)  - F(\xi(t_0))
\nn\\
&&-n^\Lambda\( \tzeta_\Lambda + \cR\(t_0^{-1}F_\Lambda - t_0 \bF_\Lambda \)- F_{\Lambda}(\xi(t_0))\)
-\frac{1}{2}\,n^\Lambda n^\Sigma F_{\Lambda \Sigma}(\xi(t_0)) - \frac{m_a}{k}\,\frac{\xi^a(t_0)}{\xi^0(t_0)} -\frac{m_0}{k}
\nn\\
&&- \frac{\cR^2}{2t_0^2\, \cS''_{\bfg,0}(t_0)}\, \Bigl(n^\Lambda \Bigl( t_0^{-1}F_\Lambda +t_0\bF_\Lambda
- \(t_0^{-1}z^\Sigma +t_0 \bz^\Sigma\) F_{\Lambda \Sigma}(\xi(t_0))\Bigr)\Bigr)^2\Biggr] .
\label{exp-saddle-I0}
\eea

The result \eqref{exp-saddle-I0} appears to be much more complicated than the standard NS5-instanton action \eqref{NS5instact}
and its physical significance is not clear to us. However, one can note that all complications come from keeping the RR fields large
so that $t_0$ remains finite and $F(\xi(t_0))$ does not reduce to $F(z)$.
Probably it is not too surprising that large RR fields lead to a weird instanton action since they couple to the self-dual 3-form
living on the world-volume of the NS5-brane, which makes the problem inherently quantum.
Below, in section \ref{subsec-smallRR} we will show that making the background RR fields small,
one reduces \eqref{exp-saddle-I0} to the expected instanton action.
Nevertheless, even without taking this additional limit, we are able to show that NS5-corrections to the hypermultiplet metric
match the quadratic structure \eqref{predictNS5} predicted by the analysis of string amplitudes.

\subsubsection{The metric and its square structure}

Using the results of the previous subsection, we conclude that at the leading order in the limit \eqref{scaling} one has
\be
\cIn{0}_\bfg\approx
- \frac{k}{p^0}\,\frac{\xi^0(t_0)}{t_0^{1+8\pi k c}}\,\frac{ e^{-S_\bfg}}{\sqrt{\I k \cS_{0}''(t_0)}}\, ,
\label{saddleI0-neg}
\ee
while all other integral functions are proportional to this one:
\begin{subequations}
\be
\cIn{\pm n}_\bfg \approx (\pm t_0)^{\mp n} \,\cIn{0}_\bfg,
\ee
\be
\cLn{n}{\bfg,\Lambda}\approx(\pm t_0)^{\mp n}\,\euF_\Lambda\, \cIn{0}_\bfg,
\qquad
\bcLn{n}{\bfg,\Lambda}\approx(\pm t_0)^{\mp n}\,\beuF_\Lambda\, \cIn{0}_\bfg,
\ee
\be
\begin{split}
\scKn{\pm n}{\Lambda\Sigma}\approx (\pm t_0)^{\mp n} \,
\euF_\Lambda\,\euF_\Sigma\,\cIn{0}_\bfg,
\qquad &
\scKn{\pm n}{\bar \Lambda \Sigma} \approx (\pm t_0)^{\mp n}\,
\beuF_\Lambda\,\euF_\Sigma\,\cIn{0}_\bfg,
\\
\scKn{\pm n}{\Lambda \bar\Sigma}\approx (\pm t_0)^{\mp n}\,
\euF_\Lambda\,\beuF_\Sigma\,\cIn{0}_\bfg,
\qquad &
\scKn{\pm n}{\bar\Lambda \bar\Sigma}\approx (\pm t_0)^{\mp n}\,
\beuF_\Lambda\,\beuF_\Sigma\,\cIn{0}_\bfg,
\end{split}
\ee
\be
\scK_\bfg \approx -\Bigl(\cR^2 K^2 +4 (z^\Lambda \euF_\Lambda)( \bz^\Sigma\beuF_\Sigma)\Bigr)\cIn{0}_\bfg,
\ee
\be
\frLn{\pm n}{\Lambda}\approx (\pm t_0)^{\mp n}\,\euF_\Lambda \de\cIn{0}_\bfg,
\qquad
\bfrLn{\pm n}{\Lambda}\approx (\pm t_0)^{\mp n}\,\beuF_\Lambda \de\cIn{0}_\bfg,
\label{smallfrG}
\ee
\label{fun-small}
\end{subequations}
where we used the functions defined in \eqref{reverseSaddle}.

As indicated above, these results are valid for positive NS5-brane charge $k$.
To get their counterparts for negative $k$, it is sufficient to replace all instances of $F(\xi)$ by $\bF(\xi)$
(including those which appear in $\euF_\Lambda$ and $\beuF_\Lambda$ \eqref{reverseSaddle}).
It is easy to check that the new saddle point $t_0^*$ is related to the old one by the antipodal map
$t_0^*=\varsigma[t_0]=-1/\bt_0$, while the instanton action satisfies $S_{-\bfg}=\bS_\bfg$.

Let us now see how the one-instanton corrected metric \eqref{ds2final2} simplifies in our limit.
We will consider only NS5-corrections given by $\frV_\bfg$ \eqref{frVtot1} with positive $k$
and extract its leading order contribution.
Then we can use the leading order results \eqref{saddleI0-neg} and \eqref{fun-small}.
In addition, there are the following simplifications:
\begin{itemize}
\item
One can drop all terms proportional to the one-loop parameter $c$ since they are always of subleading order.
\item
The terms in the first three lines of \eqref{frVtot1} are subleading compared to the rest of the expression and thus can also be dropped.
\item
The variables $r$ and $\cR$ can be exchanged (even in the perturbative part of the metric)
using the classical relation $r=\cR^2 K/4$.
\end{itemize}
As a result, the NS5 one-instanton contribution reduces to
\bea
\frV_\bfg&\approx&
-2\pi k\cIn{0}_\bfg \Biggl[
\frac{4}{\cR} \,\biggl(
(\de r)^2 -\frS^2 + r^2\,\frac{(\de K)^2}{K^2}-4r^2\cA_K^2
-\frac{2r}{K}\,|\frJ|^2
\biggr)
\nn\\
&&
+\frac{\cR}{2r^2} (z^\Lambda \euF_\Lambda) (\bz^\Sigma\beuF_\Sigma)
\biggl( (\de r)^2-\frS^2
-2r\de r\, \de\log K+4r\cA_K\frS
+r^2\,\frac{(\de K)^2}{K^2}-4r^2\cA_K^2\biggr)
\nn\\
&&
+ \frac{2}{r}\,\euF\Bigl(\de r\,\frS -2r^2\cA_K\de\log K\Bigr)
-\frac{\I\cR}{2r}\, \euF\(t_0^{-1}\frJ + t_0 \bar\frJ\) \(\frS-2r\cA_K\)
\nn\\
&&
-2\I\( \de r+r\de\log K\)\Bigl(t_0^{-1}\frJ  + t_0 \bar\frJ
-\I\cR^{-1} N^{\Lambda\Sigma} \(\beuF_\Lambda \frZ_\Sigma + \euF_\Lambda \bar\frZ_\Sigma\)
+t_0^{-1}\euF_\Lambda\de z^\Lambda + t_0 \beuF_\Lambda \de \bz^\Lambda \Bigr)
\nn\\
&&
+2\I N^{\Lambda \Sigma}\Bigl(t_0\beuF_\Lambda\frZ_\Sigma\bar\frJ +t_0^{-1}\euF_\Lambda\bar\frZ_\Sigma\frJ \Bigr)
-2\cR\Bigl(\euF_\Lambda \de z^\Lambda\bar\frJ+ \beuF_\Lambda \de \bz^\Lambda\frJ\Bigr)
\label{smallfrV}\\
&&
-\frac{1}{2r}\, \euF\(\frS-2r\cA_K\)
\Bigl(N^{\Lambda \Sigma}\(\beuF_\Sigma\frZ_\Lambda + \euF_\Sigma\bar\frZ_\Lambda \)
+\I \cR \(t_0^{-1}\euF_\Lambda \de z^\Lambda  + t_0 \beuF_\Lambda \de \bz^\Lambda \) \Bigr)
\nn\\
&&
+2\euF_\Lambda \beuF_\Sigma \Bigl(\cR^{-1}N^{\Lambda\Xi} N^{\Sigma\Theta} \frZ_\Theta \bar\frZ_\Xi-\cR \de z^\Lambda \de\bz^\Sigma\Bigr)
+2\I N^{\Lambda\Sigma} \Bigl(
t_0^{-1}\euF_\Sigma \euF_\Xi\bar\frZ_\Lambda\de z^\Xi
+t_0\beuF_\Sigma \beuF_\Xi\frZ_\Lambda \de\bz^\Xi
\Bigr))
\Biggr]
\nn\\
&-&
\[ \frac{\I}{2}\,\euF\,\de\log K
+\frac{4\I r}{\cR}\,\cA_K
-\frac{1}{\cR}\, N^{\Lambda \Sigma}\(\euF_\Lambda \bar\frZ_\Sigma - \beuF_\Lambda \frZ_\Sigma\)
+ \frac{\cR}{2r}\(z^\Lambda \euF_\Lambda \bar\frJ - \bz^\Lambda \beuF_\Lambda \frJ\) \]\de \cIn{0}_\bfg,
\nn
\eea
where we introduced another convenient notation
\be
\euF= t_0^{-1} z^\Lambda \euF_\Lambda- t_0 \bz^\Lambda \beuF_\Lambda.
\label{def-euF}
\ee

It is straightforward to verify that the expression \eqref{smallfrV} can be rewritten as
\be
\frV_\bfg \approx -\frac{\pi k}{\cR}\, \cIn{0}_\bfg \Bigl(\cA^2+\cB \de \cS_0 \Bigr),
\label{quadfrV}
\ee
where
\be
\begin{split}
\cA =&\,
2\de r + \frac{\cR\euF}{2r} \, \frS - \I\cR\(t_0^{-1}\frJ+ t_0 \bar\frJ \)
- N^{\Lambda \Sigma}\(\euF_\Lambda \bar\frZ_\Sigma+ \beuF_\Lambda \frZ_\Sigma \)
\\
&\,
+ 2r\de \log K - \cR\euF \cA_K
-\I\cR \(t_0^{-1}\euF_\Lambda \de z^\Lambda +t_0 \bar\euF_\Lambda \de \bz^\Lambda \),
\end{split}
\label{cA1form-square}
\ee
\be
\begin{split}
\cB =&\,
\frac{\cR\euF}{2r}\,\de r-2 \frS
+ \I N^{\Lambda \Sigma}\(\euF_\Lambda \bar\frZ_\Sigma- \beuF_\Lambda \frZ_\Sigma \)
- \frac{4\I}{K}\(z^\Lambda \euF_\Lambda \,\bar\frJ- \bz^\Lambda \beuF_\Lambda\, \frJ\)
\\
&\,
- \cR\(t_0^{-1}\frJ- t_0 \bar\frJ \)
+ \frac{1}{2}\,\cR\euF\, \de\log K + 4r\cA_K
- \cR \(t_0^{-1}\euF_\Lambda  \de z^\Lambda - t_0  \beuF_\Lambda \de \bz^\Lambda \)
\end{split}
\ee
and
\be
\begin{split}
\de\cS_{0}=&\,
-\[\frac{\cR\euF}{2r}\,\de r -2\frS
+\I N^{\Lambda \Sigma}\(\euF_\Lambda \bar\frZ_\Sigma- \beuF_\Lambda \frZ_\Sigma \)\]
\\
&\,
-\cR\(t_0^{-1} \frJ - t_0 \bar\frJ\)
+\frac{1}{2}\,\cR\euF\,\de\log K+4r\cA_K
-\cR \(t_0^{-1}\euF_\Lambda  \de z^\Lambda - t_0  \beuF_\Lambda \de \bz^\Lambda \) .
\end{split}
\label{fulldS0}
\ee
Note that the only dependence on the charge vector is in the overall coefficient, while $\cA$ and $\cB$ are charge independent.
One can also check that for negative $k$ the result is obtained by complex conjugation, namely, $\frV_{-\bfg}=-\bar\frV_\bfg$.
Therefore, combining \eqref{ds2final2}, \eqref{saddleI0-neg} and \eqref{quadfrV}, one finds that the full NS5-instanton correction
to the hypermultiplet metric in the small string coupling limit is given by
\be
\de s^2_{\rm NS5}\simeq
\frac{1}{4\pi r }\sum_{\bfg\; :\; k>0} \bOm_\gamma \, \frac{k^3}{p^0}\,
\Re \[\frac{\xi^0(t_0)}{t_0^{1+8\pi k c}}\,\frac{ e^{-S_\bfg}}{\sqrt{\I k \cS_{0}''(t_0)}}\,
\Bigl( \cA^2+\cB \de \cS_0\Bigr)\].
\label{resultNS5}
\ee
This is precisely the form \eqref{predictNS5} of the instanton contribution that we found from the analysis of string amplitudes.
Furthermore, comparing  \eqref{predictNS5} and \eqref{resultNS5}, we can identify (for positive $k$)
\be
\begin{split}
& \cA_\bfg=f_\bfg \,\cA+g_\bfg\, \de \cS_0,
\qquad
\cB_\bfg=\frac{1}{2\pi\I k} \(f_\bfg^2\cB-2f_\bfg g_\bfg \cA-g_\bfg^2\de \cS_0\),
\\
&\hspace{3cm}
\cN_\bfg=\frac{k^3}{8\pi r p^0}\,\frac{\xi^0(t_0)}{t_0^{1+8\pi k c}}\, \frac{f_\bfg^{-2}\, }{\sqrt{\I k \cS_{0}''(t_0)}}\, ,
\end{split}
\label{predict}
\ee
where $f_\bfg$ and $g_\bfg$ are {\it a priori} unknown functions of the moduli.
It is tempting to speculate that $g_\bfg=0$ and $f_\bfg$ is a constant.
But even keeping these functions arbitrary, the identifications \eqref{predict}
provide a large set of predictions for the amplitudes $\cA^{(\bfg)}_m(\phi)$ of one closed string vertex operator
and two fermion zero modes in the NS5-brane background.

\subsection{The limit of small RR fields}
\label{subsec-smallRR}

Unfortunately, the results \eqref{cA1form-square}-\eqref{fulldS0} providing predictions for string amplitudes
are not quite explicit due to their dependence on the solution of the saddle point equation \eqref{eqt0},
which can be solved explicitly only in simple cases of quadratic or cubic prepotential.
This is also the reason for the complicated form of the instanton action \eqref{exp-saddle-I0},
and can be traced back to the fact that
in our limit the RR fields scale as $g_s^{-1}$ so that the perturbative Darboux coordinates
$\xi_{\rm pert}^\Lambda $ scale homogeneously.
On the other hand, if the term $\cR t^{-1} z^\Lambda$ was dominating, we could expand all instances of
$F_\Lambda(\xi(t_0))$ around $\cR t^{-1} F_\Lambda(z)$ which would lead to drastic simplifications.
This approximation does hold provided we take the RR fields to be small or at least $\zeta^\Lambda,\tzeta_\Lambda\ll g_s^{-1}$.
Since, at present, calculation of string amplitudes in a non-trivial RR-background appears to be an outstanding problem,
we find it natural to consider our results in this additional limit.

We start by analyzing the saddle point equation \eqref{eqt0}.
It is easy to realize that its solution is proportional to $\zeta^\Lambda/\cR$, which is an exact result for a quadratic prepotential.
Therefore, in the limit of small RR fields, the first and third terms in the expression for $\xi_{\rm pert}^\Lambda $ \eqref{exline}
are suppressed by two orders comparing to the second term. Expanding around it, we find
\be
\begin{split}
\euF_\Lambda =&\,
-\frac{t_0}{2\cR}\, F_{\Lambda\Sigma\Theta}(z) \(\zeta^\Sigma-\cR t_0 \bz^\Sigma\)\(\zeta^\Theta-\cR t_0 \bz^\Theta\)
+O(g_s^4\zeta^5),
\\
\beuF_\Lambda =&\,  -\frac{\I\cR}{t_0}\, N_{\Lambda\Sigma} z^\Sigma -\I N_{\Lambda\Sigma}\(\zeta^\Sigma-\cR t_0 \bz^\Sigma\)
+O(g_s^2\zeta^3).
\end{split}
\ee
Substituting these expansions into \eqref{eqt0}, it is easy to solve the resulting equation on $t_0$. This gives
\be
t_0=\frac{N_{\Lambda\Sigma}\bz^\Lambda \zeta^\Sigma}{\cR N_{XY}\bz^X \bz^Y}+O((g_s\zeta)^3),
\label{t0-small}
\ee
consistently with the expectation that $t_0\sim \zeta/\cR$.

We can now perform the same expansion in the instanton action \eqref{exp-saddle-I0}.
One can observe that ignoring the next order term in \eqref{t0-small} corresponds to ignoring the terms of order $O(g_s^2\zeta^{2+n})$
with $n=0,1,2$ in contributions that scale as $g_s^{-n}$ in the limit \eqref{scaling}.
Dropping such terms and taking into account that
\be
\cS''_{0}(t_0)\approx \I \cR^2 N_{\Lambda\Sigma}\bz^\Lambda\bz^\Sigma,
\ee
it is straightforward to check that $S_\bfg$ reduces to the expected NS5-instanton action \eqref{NS5instact}
(plus a trivial constant term $4\pi k c$ as in \eqref{evalint}).
This establishes a link with the known results about these instantons and shows that they should emerge
from string amplitudes only in the approximation of small RR fields.

This double limit procedure might seem equivalent to the naive single limit where all fields are
fixed and only $r$ scales, but this is not the case.
Indeed, there are terms that survive the naive limit, but are dropped
in the limit \eqref{scaling} even before taking the RR fields to be small.
Had we scaled only $r$ from the start, these terms would have remained relevant and would change our results.
For example, we would have to change the saddle point \eqref{t0-small} by replacing $\zeta^\Lambda$
by $\zeta^\Lambda-n^\Lambda$ as in \eqref{value-tp0}.
The reason why it is the double limit rather than the naive one that should be considered is
our interest in predictions for string amplitudes.
The point is that the first limit \eqref{scaling} evaluated in the previous subsection ensures a relation between
various expansion terms and string amplitudes, while the second limit of small RR fields is supposed to be taken
already in each such term separately. In this way it simply gives the corresponding string amplitudes
in a particular region of the moduli space.
Instead, the naive limit mixes contributions from different string diagrams.
For example, in \eqref{NS5instact} the terms linear in $n^\Lambda$ originate
from $\cS_{\bfg,1}$ \eqref{Sg1} scaling as $g_s^{-1}$ and therefore are expected to capture
disk amplitudes with boundary on D-branes bound to the NS5-brane, while in the naive limit
they have a trivial scaling and are mixed with sphere contributions from $\cS_0$.

Finally, we evaluate the limit of small RR fields for the one-forms \eqref{cA1form-square}-\eqref{fulldS0},
which according to our reasoning should provide predictions for the same limit of the sphere three-point functions.
To this end, it is also useful to note that for the function defined in \eqref{def-euF} one obtains
a very simple result
\be
\euF=\I\cR K+O(g_s^3\zeta^4).
\ee
Then keeping only terms that are at most quadratic in the RR fields and using notation
\be
\hzeta^\Lambda=\zeta^\Lambda-\frac{N_{\Sigma\Theta}\bz^\Sigma \zeta^\Theta}{N_{XY}\bz^X \bz^Y}\, \bz^\Lambda,
\ee
one gets\footnote{Note that the one-forms in \eqref{limit-onrforms} are written using $t_0$ rather than its limiting value \eqref{t0-small}.
The difference manifests only in one place --- the third term in $\cB$. After expansion of this term around \eqref{t0-small},
the order $\zeta^3$ term in $t_0$ generates a contribution which is of the same order as the forth and fifth terms in $\cB$
and should be taken into account.
Since this is the only place where the correction to \eqref{t0-small} plays a role and since $\cB$ is not accessible from string amplitude
computation anyway, we did not include the correction in the main text. For completeness, we provide its expression here:
$$
-\frac{\I F_{\Lambda\Sigma\Theta}\hzeta^\Sigma\hzeta^\Theta}{6 N_{XY} \bz^X\bz^Y}
\(\zeta^\Lambda-\frac{N_{\Sigma'\Theta'}\bz^{\Sigma'} \zeta^{\Theta'}}{N_{X'Y'}\bz^{X'} \bz^{Y'}}\, \bz^\Lambda\).
$$
}
\be
\begin{split}
\hspace{-0.4cm}
\cA \approx &\,
2\de r + 2\I \frS+\I\zeta^\Lambda\frZ_\Lambda -\I\cR t_0 \bz^\Lambda(\frZ_\Lambda+\bar\frZ_\Lambda)
-\frac{\I}{2}\,F_{\Lambda\Sigma\Theta}\hzeta^\Lambda\hzeta^\Sigma \de z^\Theta
- \cR t_0 N_{\Lambda\Sigma}\hzeta^\Lambda \de \bz^\Sigma,
\\
\hspace{-0.4cm}
\cB \approx&\,
2\I \de r -2 \frS+\frac{2\cR}{t_0}\, \frJ -\zeta^\Lambda\frZ_\Lambda +\cR t_0 \bz^\Lambda(\frZ_\Lambda+\bar\frZ_\Lambda)
-\frac{1}{2}\,F_{\Lambda\Sigma\Theta}\hzeta^\Lambda\hzeta^\Sigma \de z^\Theta
- \I\cR t_0 N_{\Lambda\Sigma}\hzeta^\Lambda \de \bz^\Sigma,
\\
\hspace{-0.4cm}
\de\cS_{0}\approx &\,
-2\I \de r +2 \frS +\zeta^\Lambda\frZ_\Lambda -\cR t_0 \bz^\Lambda(\frZ_\Lambda-\bar\frZ_\Lambda)
-\frac{1}{2}\,F_{\Lambda\Sigma\Theta}\hzeta^\Lambda\hzeta^\Sigma \de z^\Theta
- \I\cR t_0 N_{\Lambda\Sigma}\hzeta^\Lambda \de \bz^\Sigma.
\end{split}
\label{limit-onrforms}
\ee
It is interesting that both $\cA$ and $\cB$ are very similar to $\de \cS_0$.
In particular, one has a very simple relation
\be
\cA\approx\I\de\cS_0-2\cR t_0 \(\I\bar\frJ+ N_{\Lambda\Sigma}\hzeta^\Lambda \de \bz^\Sigma\).
\ee

Combining the one-forms \eqref{limit-onrforms} with the identifications \eqref{predict}, one obtains predictions
for the sphere three-point functions in the NS5-background in the limit of small RR fields.
This can be viewed as one of the main results of our work.

\section{Conclusions}
\label{sec-concl}

In this paper we have computed the hypermultiplet metric of type IIA string theory compactified on a CY threefold
in the one-instanton approximation, i.e. including D-brane and NS5-brane instanton contributions
linear in the DT invariants. The resulting metric has passed two consistency checks. First, we have verified that
it is compatible with the Przanowski description of four-dimensional QK manifolds in terms of solutions of a differential equation.
Second, we have shown that in the small string coupling limit it acquires a structure that is derived from
the analysis of string amplitudes: for a fixed charge, the metric is the square of a one-form, up to contributions proportional
to the differential of the instanton action, see \eqref{sqaurestr}.
The coefficients of this one-form are expected to be sphere three-point functions of one closed
string vertex operator and two fermionic zero modes corresponding to supersymmetries broken by the NS5-brane.
We have explicitly computed this one-form and thereby provided a prediction for these three-point functions,
which takes particularly simple form \eqref{limit-onrforms} in the limit of small RR-fields.

The starting point for all these results is a set of holomorphic transition functions on the twistor space of $\cM_H$,
which encode in a concise way the instanton corrections to the metric. The transition functions for NS5-instantons
that we used in this work have been obtained in the type IIB formulation by applying S-duality to the D-instanton transition functions.
The use of type IIB results in type IIA is one of the reasons why the metric we obtained does not exhibit
manifest symplectic invariance.
One may expect that there should exist a dual formulation, obtained by some kind of Poisson resummation,
that is more adapted to symmetries of type IIA theory, simpler and allows a natural extension to multi-instanton level.
One of the motivations for our work was to find some hints for such formulation.

An interesting possibility for this is provided by the Przanowski description of the universal hypermultiplet where
we have a set of solutions \eqref{solns5} of the Przanowski differential equation which have the form of NS5-instantons and
exhibit symplectic invariance because the combination of the RR-fields in the exponential can be interpreted as
the Hesse potential associated to the prepotential \eqref{prepUHM} and evaluated on the real
symplectic vector $(\zeta,\tzeta)$. Unfortunately, it seems difficult to relate it to the solution following from our metric
as it requires solving the equation \eqref{matchNS5} which is likely not possible.
One should also recall that rigid CYs do not have mirror duals and hence the universal hypermultiplet does not have
a type IIB realization. Thus, strictly speaking, we extended our results to the case where
the starting point does not really exist. On the other hand, the metric we derived and its analysis
in the four-dimensional case rely only on the general form \eqref{5pqZ2} of the NS5 transition functions
dictated by the Peccei-Quinn symmetries and do not depend on the concrete form of the function $\Psi_{\hbfg}(\xi)$.
Therefore, if one eventually finds an alternative formulation of NS5-instantons, the corresponding transition functions
should still have the form \eqref{5pqZ2} and our results will still be valid provided one
substitutes $\Psi_{\hbfg}(\xi)$ in the definition of various integral functions
introduced in appendix \ref{ap-NS5fun} with its new counterpart.

In this respect it is worth to emphasize that although we did use the form of $\Psi_{\hbfg}(\xi)$
to derive the small string coupling limit of the instanton corrections,
our predictions for three-point string amplitudes appear to be universal and independent on a concrete formulation
of NS5-instantons. This is because the one forms \eqref{cA1form-square}-\eqref{fulldS0},
or their limit at small RR fields \eqref{limit-onrforms}, are independent of any charges.
Thus, a change of formulation, which necessarily involves a resummation over a set of charges,
will affect only the prefactor $\cN_\bfg$ in \eqref{predictNS5} related to the torus amplitude in the NS5-background.
It is an exciting problem to reproduce them from a direct worldsheet approach to string amplitudes.

\paragraph{Acknowledgements:} We are grateful to Ashoke Sen for valuable correspondence.
SA is grateful to the organizers of the program ``Black holes: bridges between number theory and holographic quantum information"
and to the Isaac Newton Institute for Mathematical Sciences where this work was finished for the kind hospitality.

\appendix

\section{NS5 transition functions}
\label{ap-HNS5}

To present the transition functions incorporating NS5-instantons obtained in \cite{Alexandrov:2010ca},
let us introduce some notations.

First, let us recall that the holomorphic prepotential describing the special K\"ahler geometry of
the K\"ahler moduli space of the mirror CY $\CYm$ is a sum of three contributions \cite{Hosono:1993qy}
\be
F=\Fcl+ F_{\rm pert} +F_{\rm ws},
\label{fullF}
\ee
where the first term is the classical prepotential
\be
\label{prepcl}
\Fcl(X)=-\frac{(X)^3}{6 X^0}+ \frac12\, A_{\Lambda\Sigma} X^\Lambda X^\Sigma,
\ee
while the other two represent perturbative $\alpha'$ and world-sheet instanton corrections, respectively.
In \eqref{prepcl}, we abbreviated $(X)^3=\kappa_{abc} X^a X^b X^c$ where
$\kappa_{abc}$ are the triple intersection numbers of $\CYm$, and
$A_{\Lambda\Sigma}$ is a real matrix satisfying
\be
\begin{split}
& \quad
A_{00}=0,
\qquad
A_{0a}= \frac{c_{2,a}}{24} ,
\\
L_a(p)\equiv &\,\frac12\, \kappa_{abc} p^b p^c-A_{ab}  p^b  \in \IZ
\qquad {\rm for}\ \forall p^a\in\IZ,
\end{split}
\label{propA}
\ee
where $c_{2,a}$ are the components of the second Chern class of $\CYm$.

The matrix $A_{\Lambda\Sigma}$ is used to define a shifted version of electric charges and Darboux coordinates
\be
q'_\Lambda = q_\Lambda -A_{\Lambda\Sigma} p^\Sigma,
\qquad
\txi'_\Lambda = \txi_\Lambda -A_{\Lambda\Sigma} \xi^\Sigma.
\label{chargeshift}
\ee
In turn, they are used to define the charges
\be
\hat q_a= q'_a + \frac12\, \kappa_{abc} \frac{p^b p^c}{p^0}\, ,
\qquad
\hat q_0 =
q'_0 +\frac{p^a q'_a}{p^0}+ \frac13\, \kappa_{abc} \frac{p^a p^b p^c}{(p^0)^2}
\label{defhatq}
\ee
that stay invariant under the so-called spectral flow transformations
generated by the monodromies around the large volume point in the K\"ahler moduli space.
Acting on the unprimed charge vector $\gamma=(p^0,p^a,q_a,q_0)$, it is realized by the matrix
\be
\rho(M_{\eps^a})
=\(\begin{array}{cccc}
1\ & 0 & 0 & 0
\\
\epsilon^a & {\delta^a}_b & 0 & 0
\\
-L_a(\epsilon)\ & -\kappa_{abc}\epsilon^c & {\delta_a}^b & \ 0
\\
\ L_0(\epsilon) & L_b(\epsilon)+2A_{bc}\epsilon^c & -\epsilon^b\ & 1
\end{array}\) ,
\label{monmat}
\ee
where $\eps^a\in \IZ$ and we introduced an integer valued function
\be
L_0(\eps)\equiv\frac16\, \kappa_{abc}\eps^a\eps^b\eps^c+\frac{1}{12}\, c_{2,a}\eps^a\in \IZ,
\label{defL0}
\ee
equal to the holomorphic Euler characteristic of the divisor specified by the vector $\eps^a$.
The properties \eqref{propA} and \eqref{defL0} ensure that the symplectic transformation \eqref{monmat} is also integer valued.
For some purposes, it is convenient to rewrite the action of the spectral flow on $\hgam=(p^a,q_a,q_0)$ as a shift
\be
\hgam\mapsto \hgam+\delta[\gamma,\eps^a],
\label{def-spflow}
\ee
where
\be
\delta[\gamma,\eps^a]=\(\begin{array}{c}
p^0 \eps^a
\\
-\kappa_{abc}p^b \eps^c-p^0 L_a(\epsilon)
\\
-q_a\eps^a +p^\Lambda L_\Lambda(\eps) +2A_{ab}p^a\eps^b
\end{array}\).
\label{def-spflowq}
\ee

Finally, we define the charge vector $\bfg=(k,p,p^a,q_a,q_0)$
where\footnote{We will take $p^0$ to be of the same sign as $k$.} $p^0=\gcd(k,p)$,
rational charges $n^0=p/k$, $n^a=p^a/k$, S-duality transformation
\be
\label{Sdualde}
g_{k,p} = \begin{pmatrix} a & b \\ - k/p^0 & p/p^0 \end{pmatrix}\in SL(2,\IZ) ,
\ee
where the integers $(a,b)$, ambiguous up to the addition of $(k/p^0,-p/p^0)$,
are chosen such that $a p + b k = p^0$,
and $\varepsilon(g_{k,p})$ equal to the logarithm of the multiplier system of the Dedekind eta function.
With these definitions, we can now write down the expression for the NS5 transition function
\cite[Eq.(5.30)]{Alexandrov:2010ca}
\be
H_{\bfg}=-\frac{\sigma_\gamma\bOm_\gamma}{(2\pi)^2}\,\frac{k}{p^0}\(\xi^0 -n^0\)
\exp\[2\pi\I \(-\frac{k}{2}\,  S_\alpha+ \frac{p^0\( k \hat q_a (\xi^a-n^a) + p^0 \hat q_0\)}{k^2(\xi^0-n^0)}
+ a\,\frac{p^0 q'_0}{k}- c_{2,a} p^a\varepsilon(g_{k,p})\)\],
\label{defHNS}
\ee
where
\be
S_\alpha= \talp+(\xi^\Lambda -2n^\Lambda)\txi'_\Lambda
+\frac{(\xi-n)^3}{3(\xi^0-n^0)}\, .
\label{defSa}
\ee
The prefactor $-\frac{k}{p^0}\(\xi^0 -n^0\)$ was absent in \cite{Alexandrov:2010ca}, but it must be included to ensure
correct transformation properties under S-duality, as has been understood in \cite{Alexandrov:2014mfa,Alexandrov:2014rca}.

Our goal is to rewrite \eqref{defHNS} in terms of type IIA variables avoiding the use of data of the mirror CY.
As a first step in this direction, we recognize the last term in \eqref{defSa} as the cubic term in the classical part
of the prepotential \eqref{prepcl}. However, it is the full prepotential $F(X)$ \eqref{fullF} that coincides by mirror symmetry
with the prepotential of the complex structure moduli space of $\CY$.
We can safely add the last two terms in \eqref{fullF} because they are mixed by S-duality with D(-1) and D1-instantons
which have been ignored in the derivation of $H_{\bfg}$ anyway.
It remains to take into account the quadratic term in \eqref{prepcl}.
Using also \eqref{chargeshift}, one finds that
\be
S_\alpha\simeq\talp+(\xi^\Lambda -2n^\Lambda)\txi_\Lambda
-2F(\xi-n)+A_{\Lambda\Sigma} n^\Lambda n^\Sigma\, .
\label{defSaIIA}
\ee

In fact, this result is not unique.
The point is that $\Fcl(X)$ is the classical part of both $F(X)$ and $\bF(X)$ because all coefficients
in \eqref{prepcl} are real. Hence another possibility is to replace $F(\xi-n)$ by $\bF(\xi-n)$ in \eqref{defSaIIA}.
Although at this point it is impossible to decide which choice is correct, this can be established by evaluating
the instanton action in the small string coupling limit (see section \ref{subsec-smallRR})
and comparing with the expected result \cite{deVroome:2006xu}.
This shows that the choice \eqref{defSaIIA} is relevant for positive $k$ and the one with $\bF$ for negative $k$.
This prescription is also consistent with the condition \eqref{reality} which relates instantons and anti-instantons.

The representation \eqref{5pqZ2} then follows upon using \eqref{defSaIIA},
our choice for the quadratic refinement $\sigma_\gamma=(-1)^{q_\Lambda p^\Lambda}$, notations \eqref{defma0}
and the fact proven in \cite[appendix D]{Alexandrov:2014rca} that under the Peccei-Quinn transformation \eqref{Heis}
the NS5 transition functions behave as
\be
T_{\eta^\Lambda,0,0}\cdot H_{\bfg}=H_{\bfg-\(0,k\eta^0,\delta[\gamma,k\eta^a/p^0]\)}.
\ee
Note that $m_\Lambda$ and $Q$ still involve various data of the mirror CY related to its K\"ahler moduli space
which seems unnatural form the type IIA point of view.
This appears to be an artifact of our hybrid approach and we leave its detailed understanding as an open issue.

\section{Procedure to derive the metric}
\label{apap-metricgen}

The general procedure to derive the metric on a QK manifold $\cM$ from the knowledge of Darboux coordinates
on its twistor space $\cZ_\cM$ was described in detail in \cite{Alexandrov:2008nk}.
Here we present it in the form adapted to the twistor description of $\cM_H$ given in section \ref{subsec-twistor}.
It consists of several steps:
\begin{enumerate}
\item
At the first step, we redefine the coordinates $\txi_\Lambda$ and $\talp$ into\footnote{The transformation \eqref{redefDC}
can be seen as a contact transformation defining Darboux coordinates in the patch around $t=0$. Then the holomorphic prepotential $F(\xi)$
has the meaning of a transition function. Together with a similar transition function to the south pole $t=\infty$ given by
$\bF(\xi)$, they provide a twistor definition of perturbative $\cM_H$ from which
the perturbative Darboux coordinates \eqref{exline} can be derived \cite{Alexandrov:2008nk}.}
\be
\begin{split}
\txii{+}_\Lambda=&\, \txi_\Lambda-F_\Lambda(\xi),
\\
\ai{+}=&\,-\hf\( \talp +\xi^\Lambda\txi_\Lambda\)+F(\xi).
\end{split}
\label{redefDC}
\ee
The advantage of the new coordinates is that their expansion around the north pole $t=0$ of $\CP$ does not
contain singular $t^{-1}$ terms.
As a result, we can write the following Laurent expansion for the set of holomorphic coordinates on
the twistor space and the contact potential \eqref{eqchip}
\be
\begin{split}
\xi^\Lambda=&\,\xi^\Lambda_{-1}t^{-1}+\xi^\Lambda_0+\xi^\Lambda_1 t+O(t^2),
\\
\txii{+}_\Lambda=&\,\txii{+}_{\Lambda,0}+\txii{+}_{\Lambda,1}\,t+O(t^2),
\\
\ai{+}=&\,4\I c\log t+\ai{+}_{0}+\ai{+}_1 t+O(t^2),
\\
\Phi=&\, \Phi_0+\Phi_1 t +O(t^2).
\end{split}
\label{expDc}
\ee

\item
In terms of the new coordinates, the contact one-form \eqref{contform} is given by
\be
\cX=\de\ai{+}+\xi^\Lambda\de\txii{+}_\Lambda.
\label{cXpole}
\ee
Substituting the expansions \eqref{expDc} into this expression
and comparing it with the canonical form $Dt$ \eqref{canform} using \eqref{relcontform}, one finds the components of
the SU(2) connection
\be
\begin{split}
p^+ &=\frac{\I}{4}\, e^{-\Phi_0}
\, \xi^\Lambda_{-1}  \de\txi^{[+]}_{\Lambda,0}\, ,
\\
p^3 &= -\frac{1}{4}\, e^{-\Phi_0} \left( \de\alpha^{[+]}_0 +
\xi^\Lambda_0   \de\txi^{[+]}_{\Lambda,0} +
\xi^\Lambda_{-1}   \de\txi^{[+]}_{\Lambda,1}  \right) -\I \Phi_1 p^+\, .
\end{split}
\label{connection}
\ee

\item
Then one computes the triplet of quaternionic 2-forms \eqref{su2curvature}.
In particular, for $\omega^3$ the formula reads
\be
\omega^3 = -4\({\rm d} p^3-2\I  p^+ \wedge p^-\).
\label{Kform}
\ee

\item
Next, one specifies the almost complex structure $J^3$ by providing a basis of (1,0) forms on $\cM$.
Such a basis was found in \cite{Alexandrov:2008nk} and, after some simplifications, it takes the following form
\be
\label{defPi}
\pi^a =\de \(\xi^a_{-1} /\xi^0_{-1} \) ,
\qquad
\tilde\pi_\Lambda= \de\txii{+}_{\Lambda,0}  ,
\qquad
\tilde\pi_\alpha = \frac{1}{2\I}\,\de\ai{+}_0 +2c \,\de\log\xi^0_{-1} .
\ee

\item
Finally, the metric is recovered as $g(X,Y) = \omega^3(X,J^3 Y)$.
To do this in practice, one should rewrite $\omega^3$, computed by \eqref{Kform} in terms of differentials of
coordinates on $\cM$, in the form which makes explicit that it is of (1,1) Dolbeault type.
Using for this purpose a basis $\pi^X$, which can be taken to be $(\pi^a,\tilde\pi_\Lambda,\tilde\pi_\alpha)$
or some its modification, the final result should look like
\be
\omega^3=2\I g_{X\bY} \pi^X\wedge \bar\pi^{Y},
\label{metom}
\ee
from which the metric readily follows as $\de s^2 =2 g_{X\bY} \pi^X \otimes \bar\pi^{Y}$.

\end{enumerate}

\section{Integrals and their differentials}
\label{ap-fun}

\subsection{Integrals of D-instanton transition functions}

The D-instanton transition functions \eqref{prepH} and all their derivatives appearing in \eqref{genDC}
depend on a single combination of Darboux coordinates determined by the D-instanton charge $\gamma$.
Therefore, it is convenient to define a function given by the integral transform \eqref{def-frI} of this combination
\be
\cJ_\gamma(t)= \opI_\gamma\[e^{-2\pi \I\(q_\Lambda \xi^\Lambda-p^\Lambda \txi_\Lambda\)}\].
\label{defcJg}
\ee
All D-instanton corrections can be expressed using this function.
We will also define its expansion coefficients around $t=0$ and $t=\infty$ as follows
\be
\begin{split}
\cJn{0}_\gamma \equiv &\, \cJ_\gamma(0)= \int_{\ellg{\gamma}}\frac{\de t}{t}\, e^{-2\pi \I \Xi_\gamma(t)},
\\
\cJn{\pm n}_\gamma \equiv &\,\frac{1}{2}\, (\pm 1)^{n-1}\left.\frac{\p^n \cJ_\gamma}{\p (t^{\pm 1})^n}\right|_{t^{\pm 1}=0}
=(\pm 1)^n  \int_{\ellg{\gamma}}\frac{\de t}{t^{1\pm n}}\, e^{-2\pi \I \Xi_\gamma(t)},
\qquad
n\ge 1,
\end{split}
\label{newfun-expand}
\ee
where we introduced
\be
\Xi_\gamma(t)\equiv q_\Lambda \xi_{\rm pert}^\Lambda-p^\Lambda \txi^{\rm pert}_\Lambda
=\Theta_\gamma+\cR\(t^{-1} Z_\gamma-t\bZ_\gamma\),
\ee
$Z_\gamma$ is the central charge function defined in \eqref{defZ} and
\be
\Theta_\gamma = q_\Lambda \zeta^\Lambda - p^\Lambda \tzeta_\Lambda.
\ee
The coefficients $\cJn{n}_\gamma$ can be evaluated in terms of modified Bessel functions
\be
\cJn{n}_\gamma=2i^{-|n|}(Z_\gamma/|\bZ_\gamma|)^{n/2}\, e^{-2\pi\I \Theta_\gamma} K_{|n|}(4\pi\cR|Z_\gamma|)
\label{cJnK}
\ee
and satisfy the following property under complex conjugation
\be
\overline{\cJn{n}_\gamma}=\cJn{-n}_{-\gamma}.
\label{bcJcJ}
\ee

Using integration by parts, it is easy to establish the relations
\begin{subequations}
\bea
Z_\gamma \cJn{1}_\gamma&=&\bZ_\gamma \cJn{-1}_\gamma,
\label{ident-JJ}
\\
\cJn{1}_\gamma &= &2\pi \I \cR\( Z_\gamma \cJn{2}_\gamma+\bZ_\gamma \cJn{0}_\gamma\),
\label{ident-J1}
\\
\cJn{-1}_\gamma &= & 2 \pi \I \cR \(Z_\gamma \cJn{0}_\gamma+\bZ_\gamma \cJn{-2}_\gamma\).
\label{ident-Jm1}
\eea
\label{relJg}
\end{subequations}
A direct evaluation of the differentials gives
\begin{subequations}
\bea
\de \cJn{0}_\gamma &= &-2\pi \I \(\cJn{0}_\gamma\de \Theta_\gamma
+\cJn{1}_\gamma\de (\cR Z_\gamma) + \cJn{-1}_\gamma\,\de (\cR \bZ_\gamma)\),
\label{ident-dJ0}
\\
\de \cJn{1}_\gamma &= &-2\pi \I \Bigl( \cJn{1}_\gamma\de \Theta_\gamma
+\cJn{2}_\gamma\de (\cR Z_\gamma) -\cJn{0}_\gamma\de (\cR \bZ_\gamma)\Bigr)
\label{ident-dJ1}
\\
\de \cJn{-1}_\gamma &= &-2\pi \I \Bigl(\cJn{-1}_\gamma\de \Theta_\gamma
-\cJn{0}_\gamma\de (\cR Z_\gamma)+\cJn{-2}_\gamma\de (\cR \bZ_\gamma)\Bigr),
\label{ident-dJm0}
\eea
\label{ident-dJ}
\end{subequations}
where $\cJn{\pm 2}_\gamma$ can be excluded using the relations \eqref{relJg}.

\subsection{Integrals of NS5-instanton transition functions}
\label{ap-NS5fun}

In contrast to the D-instanton case, NS5-corrections to different Darboux coordinates and
their differentials cannot be expressed through a single function.
Therefore, we have to introduce several functions given by the integral transform \eqref{def-frI} of
the NS5 transition functions \eqref{5pqZ2} and its derivatives. More precisely, using the shorthand notation
\be
\alpha_n=-\hf\(\tilde\alpha+(\xi^\Lambda-2n^\Lambda) \txi_\Lambda\),
\ee
we define
\be
\begin{split}
\cI_\bfg(t)=&\, \opI_\bfg\[e^{2\pi\I k \alpha_n }\Psi_{\hbfg}\],
\\
\dIu_\bfg^\Lambda(t) =&\,
\opI_\bfg\[\(\xi^\Lambda - n^\Lambda\)e^{2\pi\I k \alpha_n }\Psi_{\hbfg}\],
\\
\dId_{\bfg,\Lambda}(t)=&\, \frac{1}{2\pi\I k}\,\opI_\bfg\[e^{2\pi\I k \alpha_n }\p_{\xi^\Lambda}\Psi_{\hbfg}\],
\\
\dIuu^{\Lambda\Sigma}_{\bfg}(t)=&\,
\opI_\bfg\[\(\xi^\Lambda - n^\Lambda\)\(\xi^\Sigma - n^\Sigma\) e^{2\pi\I k \alpha_n }\Psi_{\hbfg}\],
\\
\dIdu^\Lambda_{\bfg,\Sigma}(t)=&\,
\frac{1}{2\pi \I k}\,\opI_\bfg\[\(\xi^\Lambda - n^\Lambda\) e^{2\pi\I k \alpha_n }\p_{\xi^\Sigma} \Psi_{\hbfg}\],
\\
\dIdd_{\bfg,\Lambda\Sigma}(t)=&\,
\frac{1}{(2\pi \I k)^2}\,\opI_\bfg\[e^{2\pi\I k \alpha_n }\p_{\xi^\Lambda}\p_{\xi^\Sigma} \Psi_{\hbfg}\],
\end{split}
\label{newfun}
\ee
where $\Psi_{\hbfg}$ is evaluated at $\xi^\Lambda-n^\Lambda$ as in \eqref{5pqZ2}.
It should be clear that the functions $\dIu_\bfg^\Lambda, \dId_{\bfg,\Lambda}$ are proportional to the integral transform of
the derivatives $(-\hp_{\txi_\Lambda}H_\bfg),\hp_{\xi^\Lambda}H_\bfg$, while
$\dIuu^{\Lambda\Sigma}_{\bfg}, \dIdu^\Lambda_{\bfg,\Sigma}, \dIdd_{\bfg,\Lambda\Sigma}$ correspond to the second derivatives
$\hp_{\txi_\Lambda}\hp_{\txi_\Sigma}H_\bfg,(-\hp_{\txi_\Lambda}\hp_{\xi^\Sigma}H_\bfg),\hp_{\xi^\Lambda}\hp_{\xi^\Sigma}H_\bfg$.
Since the derivatives \eqref{def-hatder} do not commute
\be
\[\hp_{\xi^\Lambda},\hp_{\txi_\Sigma}\]=2\delta_\Lambda^\Sigma\p_{\talp},
\ee
it makes sense also to introduce the function
\be
\dIud^\Lambda_{\bfg,\Sigma}(t)=
\frac{1}{2\pi \I k}\,\opI_\bfg\[ e^{2\pi\I k \alpha_n }\p_{\xi^\Sigma} \( (\xi^\Lambda - n^\Lambda)\Psi_{\hbfg}\)\]
=\dIdu^\Lambda_{\bfg,\Sigma}+\frac{1}{2\pi \I k}\,\cI_\bfg.
\ee
Similarly to \eqref{newfun-expand}, we define expansion coefficients of all these functions,
which satisfy exactly the same conjugation property as in \eqref{bcJcJ}.

We also often use the following combinations
\be
\begin{split}
\cLn{n}{\bfg,\Lambda}=&\,\tcGn{n}{\bfg,\Lambda}-F_{\Lambda\Sigma}\cGn{n}{\Sigma}_\bfg,
\qquad
\scLn{n}_\bfg=z^\Lambda\cLn{n}{\bfg,\Lambda},
\\
\bcLn{n}{\bfg,\Lambda}=&\, \tcGn{n}{\bfg,\Lambda}-\bF_{\Lambda\Sigma}\cGn{n}{\Sigma}_\bfg,
\qquad
\bscLn{n}_\bfg =\bz^\Lambda\bcLn{n}{\bfg,\Lambda},
\\
\end{split}
\label{defcL}
\ee
\be
\begin{split}
\scKn{n}{\Lambda\Sigma}=&\, \tccKn{n}{\Lambda}{\Sigma}-F_{\Lambda\Lambda'}\ccKn{n}{\Lambda'}{\Sigma}
-F_{\Sigma\Sigma'}\tcKn{n}{\Lambda}{\Sigma'}+F_{\Lambda\Lambda'}F_{\Sigma\Sigma'}\cKn{n}{\Lambda'}{\Sigma'},
\\
\scKn{n}{\Lambda\bar\Sigma}=&\, \tccKn{n}{\Lambda}{\Sigma}-F_{\Lambda\Lambda'}\ccKn{n}{\Lambda'}{\Sigma}
-\bF_{\Sigma\Sigma'}\tcKn{n}{\Lambda}{\Sigma'}+F_{\Lambda\Lambda'}\bF_{\Sigma\Sigma'}\cKn{n}{\Lambda'}{\Sigma'},
\\
\scKn{n}{\bar\Lambda\Sigma}=&\, \tccKn{n}{\Lambda}{\Sigma}-\bF_{\Lambda\Lambda'}\ccKn{n}{\Lambda'}{\Sigma}
-F_{\Sigma\Sigma'}\tcKn{n}{\Lambda}{\Sigma'}+\bF_{\Lambda\Lambda'}F_{\Sigma\Sigma'}\cKn{n}{\Lambda'}{\Sigma'},
\\
\scKn{n}{\bar\Lambda\bar\Sigma}=&\, \tccKn{n}{\Lambda}{\Sigma}-\bF_{\Lambda\Lambda'}\ccKn{n}{\Lambda'}{\Sigma}
-\bF_{\Sigma\Sigma'}\tcKn{n}{\Lambda}{\Sigma'}+\bF_{\Lambda\Lambda'}\bF_{\Sigma\Sigma'}\cKn{n}{\Lambda'}{\Sigma'},
\end{split}
\ee
\be
\scK_\bfg=z^\Lambda z^\Sigma\scKn{2}{\Lambda\Sigma}-z^\Lambda \bz^\Sigma\scKn{0}{\Lambda\bar\Sigma}
-\bz^\Lambda z^\Sigma\scKn{0}{\bar\Lambda\Sigma}+\bz^\Lambda \bz^\Sigma\scKn{-2}{\bar\Lambda\bar\Sigma},
\ee
and the one-forms
\be
\begin{split}
\frL_{\bfg,\Lambda}=&\, \de \dId_{\bfg,\Lambda}- F_{\Lambda\Sigma} \de\dIu^{\Sigma}_\bfg,
\\
\bfrL_{\bfg,\Lambda}=&\, \de \dId_{\bfg,\Lambda} - \bF_{\Lambda\Sigma} \de\dIu^{\Sigma}_\bfg,
\end{split}
\label{def-oneformNS}
\ee
so that
\be
\de \scLn{n}_\bfg=\frLn{n}{\Lambda}+\cLn{n}{\bfg,\Lambda}\de z^\Lambda,
\qquad
\de z^\Lambda\wedge \de\cLn{n}{\bfg,\Lambda}=\de z^\Lambda\wedge \frLn{n}{\Lambda}.
\label{diff-scG}
\ee
Note that the bar on the functions and the one-forms introduced above does not mean complex conjugation.
For the latter we use overline and it involves also flipping the sign of the charge and the index $n$.
For instance, we have
\be
\overline{\scLn{n}_\bfg}=\bscLn{-n}_{-\bfg}.
\label{conjrel-scLn}
\ee

Using integration by parts and the relation
\be
t\(\p_t\talp^{\rm pert}+\txi^{\rm pert}_\Lambda\p_t\xi_{\rm pert}^\Lambda-\xi_{\rm pert}^\Lambda\p_t\txi^{\rm pert}_\Lambda\)
=8\I r_{\rm pert},
\label{dtalp}
\ee
where $r_{\rm pert}$ is given in \eqref{pert-r},
one can establish the following identities
\begin{subequations}
\bea
\scLn{1}_\bfg-\bscLn{-1}_\bfg&=&-\frac{4\I r}{\cR}\, \cIn{0}_\bfg,
\label{ident-IGG}
\\
\scLn{2}_\bfg+\bscLn{0}_\bfg&=&\frac{\I}{\cR}\(\frac{1}{2\pi k}-4r\) \cIn{1}_\bfg\label{ident-IGG2},
\\
\scLn{0}_\bfg+\bscLn{-2}_\bfg&=&\frac{\I}{\cR}\(\frac{1}{2\pi k}+4r\) \cIn{-1}_\bfg,
\eea
\end{subequations}
where we dropped the index pert on $r$.
It is also straightforward to find the following differentials
\begin{subequations}
\bea
\de \cIn{0}_\bfg &=&
-2\pi\I k\biggl[2\cIn{0}_\bfg\(\frS +2r\cA_K\)
+\I N^{\Lambda\Sigma}\(\bcLn{0}{\bfg,\Lambda}\frZ_\Sigma-\cLn{0}{\bfg,\Lambda}\bar\frZ_\Sigma\)
\nn\\
&& -\cR\(\cIn{1}_\bfg \frJ+\cIn{-1}_\bfg\bar\frJ
+\cLn{1}{\bfg,\Lambda} \de z^\Lambda
+\bcLn{-1}{\bfg,\Lambda} \de \bz^\Lambda\)
-\(\scLn{1}_\bfg+\bscLn{-1}_\bfg\)\de\cR
\biggr],
\label{ident-dI0}
\\
\de\cIn{1}_\bfg &=& -2\pi\I k\biggl[
2\cIn{1}_\bfg\(\frS +2r\cA_K\)
+\I N^{\Lambda\Sigma}\(\bcLn{1}{\bfg,\Lambda}\frZ_\Sigma-\cLn{1}{\bfg,\Lambda}\bar\frZ_\Sigma\)
\nn\\
&& -\cR\(\cIn{2}_\bfg \frJ-\cIn{0}_\bfg\bar\frJ
+\cLn{2}{\bfg,\Lambda} \de z^\Lambda
-\bcLn{0}{\bfg,\Lambda} \de \bz^\Lambda\)
-\(\scLn{2}_\bfg-\bscLn{0}_\bfg\)\de\cR
\biggr],
\label{ident-dI1}
\\
\de\cIn{-1}_\bfg &=& -2\pi\I k\biggl[
2\cIn{-1}_\bfg\(\frS +2r\cA_K\)
+\I N^{\Lambda\Sigma}\(\bcLn{-1}{\bfg,\Lambda}\frZ_\Sigma-\cLn{-1}{\bfg,\Lambda}\bar\frZ_\Sigma\)
\nn\\
&& +\cR\(\cIn{0}_\bfg \frJ-\cIn{-2}_\bfg \bar\frJ
+\cLn{0}{\bfg,\Lambda} \de z^\Lambda
-\bcLn{-2}{\bfg,\Lambda} \de \bz^\Lambda\)
+\(\scLn{0}_\bfg-\bscLn{-2}_\bfg\)\de\cR
\biggr],
\label{ident-dIm1}
\eea
\label{diff-cIn}
\end{subequations}
and
\begin{subequations}
\bea
\frLn{0}{\Lambda}&=& -2\pi\I k\biggl[
2 \cLn{0}{\bfg,\Lambda}\(\frS +2r\cA_K\)
-\cR\( \cLn{1}{\bfg,\Lambda} \frJ+ \cLn{-1}{\bfg,\Lambda} \bar\frJ\)
\nn\\
&&
+\I N^{\Sigma\Xi}\(\scKn{0}{\Lambda\bar\Sigma}\,\frZ_\Xi-\scKn{0}{\Lambda\Sigma}\,\bar\frZ_\Xi\)
-\scKn{1}{\Lambda\Sigma}\de (\cR z^\Sigma)-\scKn{-1}{\Lambda\bar\Sigma}\de (\cR \bz^\Sigma)
\biggr],
\label{frG0}
\\
\bfrLn{0}{\Lambda}&=& -2\pi\I k\biggl[
2\bcLn{0}{\bfg,\Lambda}\(\frS +2r\cA_K\)
-\cR\( \bcLn{1}{\bfg,\Lambda} \frJ+ \bcLn{-1}{\bfg,\Lambda} \bar\frJ\)
\nn\\
&&
+\I N^{\Sigma\Xi}\(\scKn{0}{\bar\Lambda\bar\Sigma}\,\frZ_\Xi-\scKn{0}{\bar\Lambda\Sigma}\,\bar\frZ_\Xi\)
-\scKn{1}{\bar\Lambda\Sigma}\de (\cR z^\Sigma)-\scKn{-1}{\bar\Lambda\bar\Sigma}\de (\cR \bz^\Sigma)
\biggr],
\label{bfrG0}
\\
\frLn{1}{\Lambda}&=& -2\pi\I k\biggl[
2 \cLn{1}{\bfg,\Lambda}\(\frS +2r\cA_K\)
-\cR\( \cLn{2}{\bfg,\Lambda} \frJ- \cLn{0}{\bfg,\Lambda} \bar\frJ\)
\nn\\
&&
+\I N^{\Sigma\Xi}\(\scKn{1}{\Lambda\bar\Sigma}\,\frZ_\Xi-\scKn{1}{\Lambda\Sigma}\,\bar\frZ_\Xi\)
-\scKn{2}{\Lambda\Sigma}\de (\cR z^\Sigma)+\scKn{0}{\Lambda\bar\Sigma}\de (\cR \bz^\Sigma)
\biggr],
\label{frG2}
\\
\bfrLn{-1}{\Lambda}&=& -2\pi\I k\biggl[
2\bcLn{-1}{\bfg,\Lambda}\(\frS +2r\cA_K\)
+\cR\( \bcLn{0}{\bfg,\Lambda} \frJ- \bcLn{-2}{\bfg,\Lambda} \bar\frJ\)
\nn\\
&&
+\I N^{\Sigma\Xi}\(\scKn{-1}{\bar\Lambda\bar\Sigma}\,\frZ_\Xi-\scKn{-1}{\bar\Lambda\Sigma}\,\bar\frZ_\Xi\)
+\scKn{0}{\bar\Lambda\Sigma}\de (\cR z^\Sigma)-\scKn{-2}{\bar\Lambda\bar\Sigma}\de (\cR \bz^\Sigma)
\biggr].
\label{bfrG2}
\eea
\label{frGn}
\end{subequations}

\section{Details on the metric calculation}
\label{ap-detailsmetric}

In this appendix we provide details of the calculation of the one-instanton corrected metric \eqref{ds2final2}.
We follow the procedure described in appendix \ref{apap-metricgen} and extensively use the notations introduced
in \eqref{def-pertforms}, \eqref{connC} and in the previous appendix.

\subsection{SU(2) connection and quaternionic 2-forms}

The first step towards the metric is to evaluate the expansion coefficients near $t=0$ of $\xi^\Lambda$
and the redefined Darboux coordinates $\txii{+}_\Lambda$ and $\ai{+}$ \eqref{redefDC}.
Starting from the expressions \eqref{genDC} where the sums run over charges $\gamma$ and $\bfg$,
and using the functions defined in \eqref{defcJg}, \eqref{newfun} and \eqref{defcL},
it is straightforward to get
\begin{subequations}
\bea
\xi^\Lambda_{-1} &=& \cR z^\Lambda,
\\
\xi^\Lambda_0&=&\zeta^\Lambda+\frac{1}{8\pi^2}\sum\limits_{\gamma} \sigma_\gamma \bOm_\gamma\, p^\Lambda \cJn{0}_\gamma
- \frac{1}{8\pi^2}\sum_\bfg \bOm_\gamma\, k\,\cGn{0}{\Lambda}_\bfg,
\\
\txii{+}_{\Lambda,0}&=&\tzeta_\Lambda-F_{\Lambda\Sigma}\zeta^\Sigma
+\frac{1}{8\pi^2}\sum\limits_{\gamma} \sigma_\gamma \bOm_\gamma\, V_{\gamma\Lambda}\cJn{0}_\gamma
-\frac{1}{8\pi^2}\sum_\bfg \bOm_\gamma\, k\,\cLn{0}{\bfg,\Lambda},
\label{Lexptxi}
\\
\txii{+}_{\Lambda,1}&=&-\I\cR\bz^\Sigma N_{\Lambda\Sigma}
-\frac{1}{2\cR}\, F_{\Lambda\Sigma\Theta}\zeta^\Sigma\zeta^\Theta
+\frac{1}{8\pi^2}\sum\limits_{\gamma} \sigma_\gamma \bOm_\gamma\, \bigg[2V_{\gamma\Lambda}\cJn{1}_{\gamma}
-\frac{1}{\cR}\, F_{\Lambda\Sigma\Theta}p^\Sigma\zeta^\Theta\cJn{0}_\gamma\bigg]
\nn \\
&&
- \frac{1}{8\pi^2} \sum_\bfg \bOm_\gamma\, k\bigg[2\cLn{1}{\bfg,\Lambda}
- \frac{1}{\cR} \, F_{\Lambda \Sigma \Theta} \zeta^\Theta \cGn{0}{\Sigma}_\bfg\bigg]
,
\\
\alpi{+}_{0}&=&
-\hf\(\sigma+\zeta^\Lambda\tzeta_\Lambda-F_{\Lambda\Sigma}\zeta^\Lambda\zeta^\Sigma\)+2\I(r+c)
\nn \\
&&
+\frac{\I}{16\pi^3} \sum_{\gamma} \sigma_\gamma \bOm_\gamma
\Bigl[\( 1+2\pi\I \zeta^\Lambda V_{\gamma \Lambda}\)\cJn{0}_\gamma +2\pi\I\cR Z_\gamma \cJn{1}_{\gamma}\Bigr]
\nn \\
&&
+\frac{\I}{16\pi^3}\sum_\bfg \bOm_\gamma\[ \cIn{0}_\bfg
-2\pi\I k \zeta^\Lambda\cLn{0}{\bfg,\Lambda}
-2\pi\I k\cR\scLn{1}_\bfg\],
\eea
\label{LaurentDarboux}
\end{subequations}
where we used one more convenient notation
\be
V_{\gamma \Lambda} = q_\Lambda - F_{\Lambda \Sigma}p^\Sigma
\label{defVg}
\ee
and the definition of our coordinate $r$ \eqref{rphi} which together with \eqref{contpotconst} gives
\be
r=\frac{\cR^2}{4}\,K -c
+ \frac{\I\cR}{32\pi^2}\sum_\gamma \sigma_\gamma\bOm_\gamma \( Z_\gamma \cJn{1}_\gamma+\bZ_\gamma \cJn{-1}_\gamma\)
- \frac{\I\cR}{32\pi^2}\sum_\bfg \bOm_\gamma\, k \(\scLn{1}_\bfg+\bscLn{-1}_\bfg\).
\label{fullr}
\ee
The full contact potential \eqref{eqchip} is easily found to be
\be
\Phi(t)=\log r-\frac{1}{8\pi^2}\sum_\bfg\bOm_\gamma\, k\, \cI_\bfg(t).
\ee

Substituting the above coefficients into \eqref{connection}, using \eqref{ident-JJ}, \eqref{ident-dJ0}, \eqref{ident-IGG}
and keeping only the linear order in the instantons,
one obtains the components of the SU(2) connection
\be
\begin{split}
p^+ =&\,\frac{\I\cR}{4r}
\left[ \frJ
+\frac{1}{8\pi^2}\sum\limits_{\gamma} \sigma_\gamma \bOm_\gamma\, \Zg{}\de \cJn{0}_\gamma
-\frac{1}{8\pi^2} \sum_\bfg \bOm_\gamma\,
k \(z^\Lambda \frLn{0}{\Lambda}-\cIn{0}_\bfg \frJ\)
\right] ,
\end{split}
\ee
\bea
p^3 &=& \frac{1}{2r} \,
\Bigg\{ \frS +\(\hf\, \cR^2 K-2c\)\cA_K
+\frac{\cR}{16\pi^2}\sum_\gamma \sigma_\gamma \bOm_\gamma\(\cJn{1}_\gamma \de Z_\gamma
-\cJn{-1}_\gamma \de \bZ_\gamma  \)
\label{SU(2)connect2}
\\
&&
- \frac{\cR}{16\pi^2}\sum_\bfg \bOm_\gamma\,k \bigg[
\cIn{1}_\bfg \frJ-  \cIn{-1}_\bfg\bar\frJ
+\cLn{1}{\bfg,\Lambda} \de z^\Lambda -\bcLn{-1}{\bfg,\Lambda} \de \bz^\Lambda
+2\I\cR\de\( \frac{r}{\cR^2}\, \cIn{0}_\bfg\)
\bigg]\Bigg\}.
\nn
\eea
Plugging these results into \eqref{Kform}, one finds
\bea
\omega^3 &= & \frac{4 \de r}{r}\wedge p^3 + 8\I  p^+ \wedge p^-
+ \frac{1}{r}\,\de \zeta^\Lambda \wedge \de \tzeta_\Lambda - \frac{2\cR K}{r}\, \de \cR\wedge \cA_K
-\frac{\I\cR^2}{r}\,N_{\Lambda\Sigma} \de z^\Lambda \wedge \de \bz^\Sigma
\nn \\
&&-\frac{1}{8 \pi^2 r}\sum_{\gamma}\sigma_\gamma \bOm_\gamma\,
\Bigl(\de  \cJn{1}_{\gamma} \wedge \de (\cR Z_\gamma) - \de \cJn{-1}_\gamma \wedge \de (\cR \bZ_\gamma) \Bigr)
\label{om3} \\
&&- \frac{1}{8\pi^2r }\sum_\bfg \bOm_\gamma \,
k \bigg[ \de \tzeta_\Lambda \wedge \de \(\cR \(z^\Lambda \cIn{1}_\bfg- \bz^\Lambda \cIn{-1}_\bfg\) \)
- \de \zeta^\Lambda \wedge \de \( \cR \(F_\Lambda \cIn{1}_\bfg - \bF_\Lambda \cIn{-1}_\bfg \) \)
\nn \\
&& +\cR\(\de z^\Lambda \wedge \de\cLn{1}{\bfg,\Lambda}
- \de \bz^\Lambda \wedge \de\bcLn{-1}{\bfg,\Lambda} \)
 -\de \cR \wedge \( \cLn{1}{\bfg,\Lambda} \de z^\Lambda
-\bcLn{-1}{\bfg,\Lambda}\de\bz^\Lambda +\frac{4\I}{\cR}\,\de\( r \cIn{0}_\bfg\)\)
\bigg].
\nn
\eea

\subsection{Basis of (1,0)-forms}

The next step is to evaluate explicitly the basis of (1,0) forms \eqref{defPi}
in the almost complex structure $J^3$. This is straightforward to do given the results \eqref{LaurentDarboux}.
But the basis can actually be further simplified.
First, since $\pi^a=\de z^a$, one can drop any terms proportional to this one-form
in other basis elements. Second, it turns out to be convenient to add to $\tilde\pi_\alpha$ the term
$-\frac{\I}{2}\xi_0^\Lambda \tilde\pi_\Lambda+2r\Phi_1 p^+$.
As a result, we can choose the following one-forms to constitute our basis
\bea
&\de z^a, &
\nn\\
\cY_\Lambda &\equiv &
\frZ_\Lambda
+ \frac{1}{8\pi^2} \sum_{\gamma}\sigma_\gamma \bOm_\gamma V_{\gamma \Lambda} \de \cJn{0}_{\gamma}
- \frac{1}{8\pi^2}\sum_\bfg \bOm_\gamma\,k \,\frLn{0}{\Lambda},
\nn\\
\Sigma&\equiv & \de r + c \, \de \log (\cR^2 K) + \I\frS
+\frac{\I}{16\pi^2}\sum_{\gamma} \sigma_\gamma \bOm_\gamma \,
\Bigl(\cR\bZ_\gamma \de\cJn{-1}_{\gamma}   +\cR\cJn{1}_{\gamma} \de Z_\gamma - Z_\gamma\cJn{1}_{\gamma} \de\cR \Bigr)
\nn\\
&&
-\frac{\I}{16\pi^2}\sum_\bfg \bOm_\gamma
\,k \bigg[\cR\(\cIn{1}_\bfg \frJ-\cIn{-1}_\bfg\bar\frJ\)+\cR z^\Lambda \frLn{1}{\Lambda}-\scLn{-1}_\bfg\de\cR
\nn\\
&&\qquad
+2\cR \cLn{1}{\bfg,\Lambda} \de z^\Lambda-\cR \bcLn{-1}{\bfg,\Lambda}\de\bz^\Lambda
 +2  \cIn{0}_\bfg\(\frS+\I r \de \log K\)\bigg].
\label{holforms}
\eea
Note that with these definitions the perturbative approximation of \eqref{holforms} is expressed through the one-forms defined in
\eqref{def-pertforms}:
\be
\cY_\Lambda^{\rm pert}=\frZ_\Lambda,
\qquad
\Sigma^{\rm pert}=\frac{r+2c}{r+c}\, \de r + \I\frS.
\label{pert1forms}
\ee
It is also useful to note that
\be
p^+=\frac{\I\cR}{4r}\(1+\frac{1}{8\pi^2}\sum_\bfg \bOm_\gamma\,k\, \cIn{0}_\bfg\)z^\Lambda\cY_\Lambda,
\label{pplus}
\ee
\be
p^3 =\frac{1}{2r} \[ \Im\Sigma+\(\hf\, \cR^2K-2c\)\cA_K
- \frac{\I}{8\pi^2}\sum_\bfg \bOm_\gamma\,k \(2r \de  \cIn{0}_\bfg+\frac{r+2c}{r+c}\,\cIn{0}_\bfg\de r\)\].
\label{SU(2)connect4}
\ee

\subsection{$\omega^3$ in the holomorphic basis}

The final step is to express the quaternionic 2-form $\omega^3$ in terms of the basis \eqref{holforms}.
This is a quite tedious exercise. Therefore, we simply give the final result:
\be
\begin{split}
\omega^3 =&\, \frac{\I\Sigma\wedge \bar\Sigma}{r^2\(1+\frac{4c}{\cR^2 K}\)} -\frac{\I}{r}\(N^{\Lambda\Sigma}-\frac{\cR^2}{2r}\,z^\Lambda\bz^\Sigma\)\cY_\Lambda\wedge\bar\cY_\Sigma
+ \frac{\I \cR^2 K }{r}\, \cK_{ab}\de z^a \wedge \de \bz^b
\\
&\,
+\frac{\cR}{4\pi r}\sum_{\gamma}\sigma_\gamma \bOm_\gamma \, \omega^{\rm D}_\gamma
+\frac{\cR}{8\pi^2r }\sum_\bfg \bOm_\gamma \,k\, \omega^{\rm NS}_\bfg,
\end{split}
\label{om3final}
\ee
where
\bea
\omega^{\rm D}_\gamma &=& 2\I\cJn{0}_\gamma
\[ - \frac{1}{\cR}\,N^{\Lambda\Lambda'}\bV_{\gamma\Lambda'}\cY_\Lambda \wedge N^{\Sigma\Sigma'}V_{\gamma\Sigma'}\bcY_\Sigma
+\cR \(Z_\gamma\Sigma_K+\de Z_\gamma\)\wedge \(\bZ_\gamma\bar\Sigma_K+\de \bZ_\gamma\)
\]
\nn\\
&-&
\( Z_\gamma\cJn{1}_\gamma+\bZ_\gamma\cJn{-1}_\gamma\)
\biggl[\frac{r+c}{2 \pi r}\(
\Sigma_K\wedge \bar\Sigma_K-\frac{\p K}{K}\wedge \frac{\bar\p K}{K}\)
-\Sigma_K\wedge N^{\Lambda\Sigma} V_{\gamma\Sigma}\bcY_\Lambda
+ N^{\Lambda\Sigma}\bV_{\gamma\Sigma}\cY_\Lambda\wedge \bar\Sigma_K\biggr]
\nn\\
&+&
2N^{\Lambda\Sigma}\Bigl(\cJn{1}_\gamma \de Z_\gamma \wedge V_{\gamma\Lambda}\bcY_\Sigma
-\cJn{-1}_\gamma \bV_{\gamma\Lambda}\cY_\Sigma\wedge \de \bZ_\gamma \Bigr),
\eea
\bea
\omega^{\rm NS}_\bfg &=&
\frac{1}{2r}\(\scLn{1}_\bfg+\bscLn{-1}_\bfg\) \[
(r+2c)\Sigma_K \wedge\bar\Sigma_K
-2(r+c)\, \frac{\p K}{K}\wedge \frac{\bar\p K}{K}\]
-
\Sigma_K\wedge \bcLn{-1}{\bfg,\Lambda}\de\bz^\Lambda
-\cLn{1}{\bfg,\Lambda} \de z^\Lambda\wedge \bar\Sigma_K
\nn\\
&-&
2\pi\I k \Biggl\{
\frac{2}{\cR} \,\cIn{0}_\bfg\(\Sigma+\frac{2r}{K}\, \p K\)\wedge \(\bar\Sigma+\frac{2r}{K}\, \bar\p K\)
+\frac{2\I (r+c)}{K(r+2c)} \(\scLn{1}_\bfg+\bscLn{-1}_\bfg\)
\Bigl(\p K\wedge \bar\Sigma-\Sigma \wedge \bar\p K\Bigr)
\nn\\
&&
+\cR\Bigl[\(\scLn{2}_\bfg-\bscLn{0}_\bfg\)z^\Lambda\cY_\Lambda \wedge \bar\Sigma_K
-\(\scLn{0}_\bfg-\bscLn{-2}_\bfg\)\Sigma_K\wedge\bz^\Lambda\bcY_\Lambda\Bigr]
\nn\\
&&
-2\[\I\cIn{1}_\bfg z^\Lambda \cY_\Lambda+\frac{N^{\Lambda\Sigma}}{\cR}\bcLn{0}{\bfg,\Lambda}\cY_\Sigma
+\I\cLn{1}{\bfg,\Lambda}\de z^\Lambda\]
\wedge \(\bar\Sigma+\frac{2r}{K}\, \bar\p K\)
\nn\\
&&
+2 \(\Sigma+\frac{2r}{K}\, \p K\)\wedge
\[\I\cIn{-1}_\bfg \bz^\Lambda \bcY_\Lambda-\frac{N^{\Lambda\Sigma}}{\cR}\cLn{0}{\bfg,\Lambda}\bcY_\Sigma
+\I\bcLn{-1}{\bfg,\Lambda}\de \bz^\Lambda\]
\nn\\
&&
+2\Bigl[ z^\Xi\cY_\Xi \wedge\Bigl(\I N^{\Lambda\Sigma}\cLn{1}{\bfg,\Lambda}\bcY_\Sigma-\cR\bcLn{0}{\bfg,\Lambda}\de\bz^\Lambda\Bigr)
-\Bigl(\I N^{\Lambda\Sigma}\bcLn{-1}{\bfg,\Lambda}\cY_\Sigma+\cR\cLn{0}{\bfg,\Lambda}\de z^\Lambda\Bigr)\wedge \bz^\Xi\bcY_\Xi\Bigr]
\nn\\
&&
-2\cR\cIn{0}_\bfg z^\Lambda\cY_\Lambda \wedge \bz^\Sigma\bcY_\Sigma
+\frac{\cR}{2}\,\scK_\bfg\, \Sigma_K\wedge \bar\Sigma_K
\nn\\
&&
+ \Sigma_K\wedge
\[\I N^{\Lambda\Sigma}\(z^\Xi\scKn{1}{\Xi\Sigma}+\bz^\Xi\scKn{-1}{\bar\Xi\Sigma}\)\bcY_\Lambda
-\cR \(z^\Lambda\scKn{0}{\Lambda\bar\Sigma}-\bz^\Lambda\scKn{-2}{\bar\Lambda\bar\Sigma}\)\de \bz^\Sigma\]
\nn\\
&&
-\[ \I N^{\Lambda\Sigma}\(z^\Xi\scKn{1}{\Xi\bar\Sigma}+\bz^\Xi\scKn{-1}{\bar\Xi\bar\Sigma}\)\cY_\Lambda
-\cR \(z^\Lambda\scKn{2}{\Lambda\Sigma}-\bz^\Lambda\scKn{0}{\bar\Lambda\Sigma}\)\de z^\Sigma
\]\wedge \bar\Sigma_K
\nn\\
&&
+\(\scKn{0}{\Lambda\bar\Sigma}+\scKn{0}{\bar\Sigma\Lambda}\)
\(\frac{1}{\cR}\,N^{\Lambda\Xi} N^{\Sigma\Theta} \cY_\Theta\wedge \bcY_\Xi-\cR\de z^\Lambda\wedge \de\bz^\Sigma\)
\nn\\
&&
+2\I N^{\Lambda\Sigma}\(\scKn{1}{\Sigma\Xi}\de z^\Xi\wedge \bcY_\Lambda-\scKn{-1}{\bar\Sigma\bar\Xi}\cY_\Lambda\wedge \de\bz^\Xi\)
\Biggr\},
\label{om3-final-NS}
\eea
and we used the shorthand notations \eqref{defVg} and
\be
\Sigma_K=\frac{\Sigma}{2(r+2c)}-\frac{\p K}{K}\, .
\ee
Importantly, \eqref{om3final} is manifestly a (1,1)-form.
To check that it coincides with the original expression \eqref{om3},
it is much easier to start from the end: one should substitute
the expressions \eqref{holforms} for the (1,0)-forms $\cY_\Lambda$ and $\Sigma$ into \eqref{om3final},
expand up to linear order in DT invariants, and compare with \eqref{om3} after using there
the results for the differentials of the integral functions \eqref{ident-dJ}, \eqref{diff-scG}, \eqref{diff-cIn} and \eqref{frGn}.
Note that in our approximation, one can replace $\cY_\Lambda$ and $\Sigma$ appearing in
$\omega^{\rm D}_\gamma$ and $\omega^{\rm NS}_\bfg$ by their simple perturbative expressions \eqref{pert1forms}.

The metric follows from \eqref{om3final} by replacing all wedge products by tensor products and multiplying
the resulting expression by $-\I$. Then, to extract the one-instanton approximation, one should again substitute
$\cY_\Lambda$ and $\Sigma$ from \eqref{holforms}, in all perturbative terms express all instances
of $\cR$ in terms of $r$ by means of \eqref{fullr}, and keep only terms linear in DT invariants.
In doing this, one can use the following identities valid up to the first instanton order
\bea
|\Sigma|^2 &=&
\frac{(r+2c)^2}{(r+c)^2}\, (\de r)^2 +\frS^2
+\frac{\cR}{8\pi^2}\sum_{\gamma} \sigma_\gamma \bOm_\gamma \Biggl\{
\Bigl(\cJn{1}_{\gamma}\de Z_\gamma -\cJn{-1}_{\gamma}\de\bZ_\gamma\Bigr)\frS
\nn\\
&&
+\frac{\I r(r+2c)}{4\pi(r+c)^2}
\biggl[2 \de \Bigl(Z_\gamma \cJn{1}_{\gamma}+\bZ_\gamma \cJn{-1}_{\gamma}\Bigr)
- \Bigl(Z_\gamma\cJn{1}_{\gamma} +\bZ_\gamma\cJn{-1}_{\gamma}\Bigr) \(\frac{\de r}{r+c}-\frac{r+2c}{r}\,\frac{\de K}{K}\)
\biggr]\de r
\Biggr\}
\nn\\
&-&
\frac{\cR}{8\pi^2}\sum_\bfg \bOm_\gamma\,k\,\Biggl\{
\frac{\I r(r+2c)}{4(r+c)^2}\, \biggl[ 2\Bigl(\de \scLn{1}_\bfg+\de \bscLn{-1}_\bfg\Bigr)
-\(\scLn{1}_\bfg+\bscLn{-1}_\bfg\)\(\frac{\de r}{r+c}-\frac{r+2c}{r}\frac{\de K}{K}\)
\biggr]\de r
\nn\\
&&
+\bigg[\frac{2\I}{\cR}\, \cIn{0}_\bfg\(\frac{c\de r}{r+c}+r\de\log K\)- \frac{2\I r}{\cR}\,\de \cIn{0}_\bfg
+\(\cLn{1}{\bfg,\Lambda} \de z^\Lambda-\bcLn{1}{\bfg,\Lambda} \de \bz^\Lambda\)
+ \cIn{1}_\bfg\, \frJ-\cIn{-1}_\bfg\, \bar\frJ\bigg]\frS
\Biggr\},
\nn
\eea
\bea
N^{\Lambda\Sigma}\cY_\Lambda\bar\cY_\Sigma
&=&
N^{\Lambda\Sigma}\frZ_\Lambda \bar\frZ_\Sigma
-\frac{\I}{8\pi^2} \sum_{\gamma}\sigma_\gamma \bOm_\gamma \,\de\Theta_\gamma\,\de \cJn{0}_{\gamma}
-\frac{\I}{8\pi^2}\sum_\bfg\bOm_\gamma\,k
\(\de\cGn{0}{\Lambda}_\bfg\de\tzeta_\Lambda-\de\tcGn{0}{\bfg,\Lambda} \de \zeta^\Lambda\),
\nn\\
\frac{\cR^2}{2r^2}\,\left|z^\Lambda\cY_\Lambda\right|^2 &=&
\(\frac{2(r+c)}{r^2 K}-\frac{2\rin}{r^2 K}\)|\frJ|^2
- \frac{\I(r+c)}{2\pi^2 r^2 K} \sum_{\gamma}\sigma_\gamma \bOm_\gamma
\Im(\bZ_\gamma \frJ)\,\de \cJn{0}_{\gamma}
\\
&&
- \frac{r+c}{2\pi^2 r^2 K}\sum_\bfg \bOm_\gamma\,
k \, \Re\(z^\Lambda\frLn{0}{\Lambda}\,\bar\frJ\),
\nn
\eea
where $\rin$ denotes the instanton terms on the r.h.s. of \eqref{fullr}.
The final result turns out to be the one-instanton corrected metric given in \eqref{ds2final2}.

\section{Details on the universal hypermultiplet case}
\label{ap-UHM}

Here we provide some useful formulae concerning the universal hypermultiplet and its Przanowski description.

First, the relation between the derivatives with respect to the real variables
and the complex Przanowski coordinates \eqref{pert-zz} is given by
\be
\begin{split}
\p_1=&\, -\frac{r+c}{2(r+2c)}\, \p_r+2\I\p_\sigma,
\\
\p_2=&\, \hp_2 + \zeta \p_1,
\qquad
\hp_2= \frac{1}{\tau_2}\(\btau(\p_{\tzeta}+\zeta\p_\sigma)-(\p_\zeta-\tzeta\p_\sigma)\).
\end{split}
\label{der12x}
\ee
Note that we have the relations
\be
\begin{split}
\p_{2}\p_{\bar1}=&\, \hp_2 \p_{\bar 1}+  \zeta \p_{1}\p_{\bar 1},
\\
\p_{1}\p_{\bar2}=&\,\p_{1} \hp_{\bar 2} +\zeta \p_{1}\p_{\bar1} ,
\\
\p_2 \p_{\bar 2} =&\, \hp_2 \hp_{\bar 2}
+ \zeta (\p_1 \hp_{\bar 2} + \hp_2 \p_{\bar 1})- \zeta^2 \p_{1}\p_{\bar 1}+ \frac{1}{\tau_2}\, \p_{\bar 1}
\end{split}
\label{relderhat}
\ee
and the shifted derivatives $\hp_2$ are non-commutative, namely $\hp_2 \hp_{\bar 2} \neq \hp_{\bar 2} \hp_2$.
The relation \eqref{der12x} leads to the following derivatives of the perturbative Przanowski potential \eqref{pert-h0}
\be
\hpert_{1}=\frac{1}{2r}\, ,
\qquad
\hpert_{2}= \zeta \hpert_1,
\label{firstderh}
\ee
\be
\hpert_{11}=
\frac{r+c}{4r^2(r+2c)}\, ,
\qquad
\hpert_{12}=\zeta \hpert_{11},
\qquad
\hpert_{22}=\zeta^2 \hpert_{11}
-\frac{1}{2\tau_2 r}\, .
\label{secondderh}
\ee
\be
\begin{split}
&\,\hpert_{111}=
\frac{(r+c)(2r^2+5cr+4c^2)}{8r^3(r+2c)^3}\, ,
\qquad
\hpert_{112}=\zeta \hpert_{111},
\\
\hpert_{122}=&\, \zeta^2 \hpert_{111}
-\frac{r+c}{4\tau_2 r^2(r+2c)}\, ,
\qquad
\hpert_{222}=\zeta^3 \hpert_{111}
-\frac{3\zeta(r+c)}{4\tau_2 r^2(r+2c)}\, .
\end{split}
\label{thirdderh}
\ee

Given that for the quadratic prepotential
\be
K=N_{00}=2\tau_2,
\ee
it is straightforward to compute the Laplace-Beltrami operator using its standard definition in terms of the metric
which we take to be the specification of the perturbative metric \eqref{1lmetric} to the four-dimensional case.
The result can be written as\footnote{In the third
term all derivatives are supposed to be on the right and not act on the coefficients.}
\be
\Delta=
\frac{r^2}{r+2c}\[
(r+c)\p_r^2+\frac{16(r+2c)^2}{r+c}\, \p_{\sigma}^2
+\frac{2}{\tau_2}\left|(\tzeta+\tau\zeta)\p_\sigma+\tau\p_{\tzeta}-\p_\zeta\right|^2
-\frac{r+2c}{r}\, \p_r
\].
\label{Laplacereal}
\ee
This formula can also be obtained by linearizing the Przanowski differential equation \eqref{Prz-master-equation}
and writing the result as in \eqref{linPrz}.

Let us now compute the deformation of the metric \eqref{deltamet} induced by the instanton corrections
to the Przanowski potential \eqref{ourdeltah} and the complex coordinates \eqref{Prz-delta-cplxcoord},
and hence given by the two contributions whose general expressions can be found in
\eqref{linPrzmet} and \eqref{coord-Przmt}.
To compute them explicitly, we will use that the differentials of the non-deformed complex coordinates \eqref{pert-zz}
are given by
\be
\label{pert-dz}
\de z_0^1 = -\I \frS - \frac{r+2c}{r+c}\,\de r - \zeta \de z_0^2,
\qquad
\de z_0^2 = \frac{\I}{2}\, \frJ.
\ee
Substituting this into \eqref{linPrzmet} and using the relations \eqref{relderhat}, one can show that
the first contribution takes the form
\be
\begin{split}
\label{linPrzmet-simpl}
\delta_h \de s^2 =&\,
4 \delta h_{1\bar1} \left|\frac{r+2c}{r+c}\de r+ \I \frS\right|^2 -2  \(\delta h_{\hat 2 \bar 1} \frJ + \delta h_{1\hat{\bar 2}} \bar \frJ \)\frS
\\
&\,
- \frac{2\I(r+2c)}{r+c} \(\delta h_{\hat 2 \bar 1} \frJ - \delta h_{1\hat{\bar 2}} \bar \frJ \)\de r
+|\frJ|^2\(  \delta h_{\hat 2\hat{\bar 2}} - \frac{1}{\tau_2} \delta h_{\bar 1}+ \frac{r+c}{\tau_2 r^2}\, \delta h\),
\end{split}
\ee
where the subscript $\hat 2$ denotes the shifted derivative $\hp_2$.
In a similar way we can evaluate the second contribution \eqref{coord-Przmt}.
In view of \eqref{Prz-delta-cplxcoord}, it is natural to write it in terms of
$\delta \hz^1\equiv \delta z^1 + \zeta \delta z^2$.
As a result, one finds
\be
\begin{split}
\label{coord-Przmet-simpl}
\delta_z \de s^2 =&\,
8\Re(\delta \hz^1 )\(\hpert_{111}\left|\frac{r+2c}{r+c}\,\de r+ \I \frS\right|^2
+\frac{r+4c}{4\tau_2 r}\,\hpert_{11} |\frJ|^2\)
+\frac{8\hpert_{11}}{\tau_2} \frS \Re (\delta z^2 \bar \frJ)
\\
&
-\frac{2}{r^2}\(\de r \Re (\de \delta \hz^1)+ \frac{r+c}{r+2c}\,\frS\Im(\de \delta \hz^1)\)
+ \frac{2(r+2c)}{\tau_2 r^2}\, \Im \(\bar \frJ\,\de \delta z^2\).
\end{split}
\ee

We still need to compute the derivatives $\delta h_{\alpha \bar \beta}$ and the differentials
$\de \delta \hz^1$ and $\de \delta z^2$.
This requires computing the action of the derivatives $\p_1$ and $\hp_2$
on the integral functions introduced in appendix \ref{ap-fun}. Here we provide some of them:
\be
\begin{split}
\label{derJgz}
\p_1\cJn{0}_\gamma=&\,
\frac{\pi \I \cR}{2(r+2c)} \( (q+\tau p)\cJn{1}_\gamma+(q+\btau p)\cJn{-1}_\gamma\),
\\
\p_1\cJn{\pm n}_\gamma=&\,
\pm \frac{\pi \I \cR}{2(r+2c)}\( (q+\tau p)\cJn{\pm n+1}_\gamma-(q+\btau p)\cJn{\pm n-1}_\gamma\),
\qquad n\ge 1,
\\
\hp_2\cJn{\pm n}_\gamma=&\, \frac{2\pi\I}{\tau_2}\, (q+\btau p)\cJn{\pm n}_\gamma,
\qquad n\ge 0,
\end{split}
\ee
\be
\begin{split}
\label{derJgzz}
\p_1\p_{\bar 1}\cJn{0}_\gamma=&\,
\frac{\pi\I \cR (r+c)}{4(r+2c)^3} \( (q+\tau p)\cJn{1}_\gamma+(q+\btau p)\cJn{-1}_\gamma\)
+\frac{2\pi^2(r+c)}{\tau_2(r+2c)^2}\, |q+\tau p|^2 \cJn{0}_\gamma,
\\
\p_1\hp_{\bar 2}\cJn{0}_\gamma=&\,
-\frac{\pi^2\cR}{\tau_2(r+2c)} \( (q+\tau p)^2\cJn{1}_\gamma+|q+\btau p|^2\cJn{-1}_\gamma\)
\\
\hp_2\hp_{\bar 2} \cJn{0}_\gamma=&\,-\frac{4\pi^2}{\tau_2^2}\, |q+\tau p|^2  \cJn{0}_\gamma,
\end{split}
\ee
\be
\begin{split}
\label{derIz}
\p_1\cIn{0}_\bfg=&\, 2\pi k\, \cIn{0}_\bfg
-\frac{\pi\I k\cR}{2(r+2c)}\(\scLn{1}_\bfg+\bscLn{-1}_\bfg\),
\\
\hp_2 \cIn{0}_\bfg=&\, -\frac{2\pi\I k}{\tau_2}\, \bscLn{0}_\bfg
+4\pi k\cR\,\cIn{1}_\bfg,
\end{split}
\ee
\be
\begin{split}
\p_1\scLn{0}_\bfg=&\, 2\pi k\, \scLn{0}_\bfg
-\frac{\pi\I k\cR}{2(r+2c)} \(\scKn{1}{00}+\scKn{-1}{0\bar 0}\),
\\
\hp_2 \scLn{0}_\bfg=&\,  -\frac{2\pi\I k}{\tau_2}\, \scKn{0}{0\bar 0}
+4\pi k\cR\,\scLn{1}_\bfg,
\end{split}
\label{derscLn0}
\ee
\be
\begin{split}
\p_1\scLn{1}_\bfg=&\, 2\pi k\, \scLn{1}_\bfg
-\frac{\pi\I k\cR}{2(r+2c)} \(\scKn{2}{00}-\scKn{0}{0\bar 0}\),
\\
\hp_2\scLn{1}_\bfg=&\,  -\frac{2\pi\I k}{\tau_2}\, \scKn{1}{0\bar 0}
+4\pi k\cR\,\scLn{2}_\bfg,
\end{split}
\label{derscLn1}
\ee
\be
\begin{split}
\label{derIzz}
\p_1\p_{\bar 1}\cIn{0}_\gamma=&\, -4\pi^2 k^2\, \cIn{0}_\gamma
-\frac{\pi\I k\cR r}{8(r+2c)^3}\(\scLn{1}_\bfg+\bscLn{-1}_\bfg\)
-\frac{\pi^2 k^2(r+c)}{2\tau_2(r+2c)^2}\,\scK_\bfg\, ,
\\
\p_1\hp_{\bar 2}\cIn{0}_\gamma=&\,
-\frac{4\I\pi^2 k^2}{\tau_2(r+2c)}\((2r+3c)\scLn{0}_\bfg-(r+c)\bscLn{-2}_\bfg\)
\\
&\,
-\frac{\pi^2 k^2\cR }{\tau_2(r+2c)}\(\scKn{1}{00}+\scKn{-1}{\bar 00}\)
-8\pi^2k^2 \cR \cIn{-1}_\bfg,
\\
\hp_2\hp_{\bar 2} \cIn{0}_\gamma=&\,
-\frac{2\pi^2 k^2 }{\tau_2^2} \(\scKn{0}{0\bar 0} + \scKn{0}{\bar00}\)
-\frac{2\pi k }{\tau_2}\(1-16\pi k c\)\cIn{0}_\bfg.
\end{split}
\ee
The complex conjugate derivatives can be obtained using \eqref{bcJcJ} and similar relations
for the functions associated with NS5-instantons like, e.g., \eqref{conjrel-scLn}.
Besides, one can obtain derivatives of $\bscLn{0}_\bfg$  from \eqref{derscLn0}
by putting bar on functions $\scLn{n}_\bfg$ and on the first index of functions $\scK^{(n)}_\bfg$,
and derivatives of $\bscLn{-1}_\bfg$ from \eqref{derscLn1} by doing the same plus lowering all upper indices by 2 and
flipping the sign of the second terms on the r.h.s.

It is a straightforward although a bit tedious exercise to substitute \eqref{ourdeltah} and \eqref{Prz-delta-cplxcoord} into
\eqref{linPrzmet-simpl} and \eqref{coord-Przmet-simpl}, respectively, and use the above results for derivatives
of the integral functions to compute them explicitly. Then combining the two contributions, one finally obtains
\be
\delta \de s^2=
\frac{\cR}{8 \pi^2 r}\sum_{\gamma}\sigma_\gamma \bOm_\gamma \, \frD_\gamma
+\frac{\cR}{8\pi^2r }\sum_{\bfg} \bOm_\gamma \,k\, \frV_{\bfg},
\ee
where
\bea
\frD_\gamma &=&
\frac{\pi}{\cR}\, \cJn{0}_\gamma \[
\cR^2|Z_\gamma|^2 \( \frac{\Im \hat \Sigma ^2}{(r+2c)^2}-\frac{\de r^2}{(r+c)^2}\)
-\frac{3r+4c}{r}\,\de \Theta_\gamma^2 -4\cC_\gamma^2 \]
\label{frDtotUHM}\\
&&
+ \(Z_\gamma \cJn{1}_\gamma + \bZ_\gamma \cJn{-1}_\gamma\) \Bigg[
\frac{\I \frS^2}{4(r+2c)^2}- \frac{\I \de r^2}{4(r+c)^2} - \frac{\I |\frJ|^2}{4r\tau_2}
+ \frac{2\pi \cC_\gamma\frS}{r+2c}- \frac{\pi (r+2c)}{r(r+c)}\,\de r \de \Theta_\gamma\Bigg],
\nn
\eea
and
\bea
\label{frVtotUHM}
\frV_\bfg &=&\,
\(\scLn{1}_\bfg + \bscLn{-1}_\bfg\) \[\frac{\I}{8}\(\frac{\de r^2}{(r+c)^2} -  \frac{\frS^2}{(r+2c)^2}\)+\frac{\I|\frJ|^2}{4r\tau_2} \]
\nn\\
&&\,
- \frac{2(r+c)}{r(r+2c)}\, \frS \[\frac{\I c}{\cR}\,\cIn{0}_\bfg \,\frac{\de r}{r+c}+ \Re (\cIn{1}_\bfg \frJ) \]
- \frac{\I (r+2c)}{2r(r+c)}\Re ( \de \scLn{0}_\bfg \bar \frJ)
\nn\\
&&\,
- 2\pi k \Bigg \lbrace
\frac{2}{\cR}\,\cIn{0}_\bfg \[ \frac{(r+2c)^2}{(r+c)^2}\,\de r^2 - \frac{r\frS ^2}{r+2c} - \frac{2c}{\tau_2} \,|\frJ|^2\]\\
&&\,
\quad
+ \frac{\frS}{r+2c}\[\cR\Im \(\scLn{2}_\bfg\frJ \)
-\frac{\cR(r+2c)}{r+c}\,\Im \(\bscLn{0}_\bfg \frJ \)
-4c \Re\(\cIn{1}_\bfg\frJ\)\]
\nn\\
&&\,
\quad
+\frac{r+2c}{r+c}\,\de r\[ 4\Im \(\cIn{1}_\bfg\frJ\)
-\frac{\cR}{r+c} \Re \(\bscLn{0}_\bfg\frJ\)\]
+ \frac{(2r+3c)\frS\de r }{(r+c)(r+2c)} \(\scLn{1}_\bfg + \bscLn{-1}_\bfg\)
\nn\\
&&\,
\quad
+\frac{\cR}{8}\scK_\bfg \[ \frac{\frS^2}{(r+2c)^2}-\frac{\de r^2}{(r+c)^2}\]
- \frac{\frS}{2\tau_2(r+2c)} \Re \bigl[\bigl(\scKn{1}{0\bar0} + \scKn{-1}{\bar0\bar0}\bigr)\frJ\bigr]
+ \Re\bigl(\scKn{0}{0\bar0}\bigr)\,\frac{|\frJ|^2}{2\tau_2^2 \cR }
\Bigg \rbrace.
\nn
\eea
It is easy to check that these expressions for $\frD_\bfg$ and $\frV_\bfg$ coincide with \eqref{frDtot2}
and \eqref{frVtot1}, respectively, specified to the one-modulus case. This confirms that the instanton corrected metric
computed in section \ref{sec-metric} is consistent with the Przanowski description.

\section{One-loop effects at small $g_s$}
\label{ap-Sinst}

Let us consider the integral
\be
\int_{\ell_\bfg}\frac{\de t}{t^{n+1}} \, e^{-2\pi\I k\cS_\bfg(t)},
\label{inttn}
\ee
which is a particular case of the integral in \eqref{intft}.
We are interested in its leading behavior in the limit $r\to\infty$ keeping all other fields
$z^a,\zeta^\Lambda,\tzeta_\Lambda$ and $\sigma$ fixed.
Of course, in this limit we also have $\cR\sim r^{1/2}$.
Since the action $\cS_\bfg(t)$ scales as $r$, one can hope to apply the saddle point
approximation. It is easy to see that, for positive $k$, the saddle point scales itself as $r^{-1/2}$ (cf. \eqref{t0-small}).
Therefore, we can redefine the integration variable as $t=t'/\cR$ and expand $\cS_\bfg(t)$ in inverse powers of $\cR$.
This gives
\be
\begin{split}
\cS_\bfg(t')=&\,
- \frac{\I}{2}\,  \cR^2 K+\hf \( \sigma + \tzeta_\Lambda \zeta^\Lambda- 2n^\Lambda \tzeta_\Lambda\)
- \hf\, F_{\Lambda \Sigma}(\zeta^\Lambda-n^\Lambda)(\zeta^\Sigma-n^\Sigma)
- \frac{m_\Lambda}{k}\,z^\Lambda
\\
&\,
-\I t'  N_{\Lambda\Sigma}\bz^\Lambda(\zeta^\Sigma-n^\Sigma) +\frac{\I}{2}\, t'^2 N_{\Lambda\Sigma} \bz^\Lambda\bz^\Sigma
-4\I c  \log (t'/\cR) + O(\cR^{-1}).
\end{split}
\label{approx-cS}
\ee
If one sets $c=0$ thereby ignoring the logarithmic term, the saddle point is found to be
\be
t'_0=\frac{N_{\Lambda\Sigma}\bz^\Lambda(\zeta^\Sigma-n^\Sigma)}{N_{XY}\bz^X\bz^Y}
\label{value-tp0}
\ee
and $2\pi\I k \cS_\bfg(t_0')=S_\bfg^{(0)}$ reproducing the instanton action \eqref{NS5instact} found in \cite{Alexandrov:2010ca}.

If however one retains the one-loop parameter $c$ non-vanishing, the saddle point approximation appears to break down.
While the initial integral \eqref{inttn} is still localized at small $t$ so that the expansion \eqref{approx-cS}
remains justified, it is not true anymore that the leading contribution
comes from a Gaussian integral near $t'=t'_0$. Instead, one finds that it is given by
\be
\cR^{n+8\pi k c}\, \cD_{-n-8\pi k c}\(-\pi k N_{\Lambda\Sigma} \bz^\Lambda\bz^\Sigma ,t'_0\)e^{-(S_\bfg^{(0)}+4\pi k c)},
\label{evalint}
\ee
where
\be
\cD_\nu(a,b)=\int_{-\infty}^\infty \de s\, s^{\nu-1} \, e^{-a(s-b)^2}
\ee
and we assume that the logarithmic branch cut is in the positive half plane, while the integration contour avoids
the singularity at $s=0$ from below. Since the factor given by the function $\cD_\nu$ does not contain the scaling parameter,
it cannot be simplified anymore. In fact, it can be evaluated in terms of the parabolic cylinder function $D_\nu(z)$
using the formula \cite[3.462.3]{GradRyzh}
\be
\int_{-\infty}^{\infty}\de s\, (\I s)^{\nu}\, e^{-a s^2 - \I q s} = 2^{-\frac{\nu}{2}} \sqrt{\pi}\, a^{-\frac{\nu+1}{2}} \,
e^{-\frac{q^2}{8a}} D_{\nu}\(\frac{q}{\sqrt{2a}}\),
\label{Gradstein-int}
\ee
valid for $\Re a > 0$ and $\Re \nu > -1$.
The last condition can be dropped once we do an analytic continuation which precisely corresponds to the deformation
of the contour away from the singularity at the origin.
As a result, we obtain
\be
\cD_\nu(a,b)=(-2)^{-\frac{\nu-1}{2}} \sqrt{\pi}\, a^{-\frac{\nu}{2}}\,
e^{-\hf\, a b^2} D_{\nu-1}\(\I \sqrt{2a}\,b\).
\ee
Since for large arguments $D_\nu(z)$ behaves like $z^\nu e^{-z^2/4}$ times a polynomial in $1/z$,
in the function $\cD_\nu(a,b)$ the exponential factors cancel so that it scales as $b^\nu$.
Therefore, it does not affect the instanton action in \eqref{evalint}.

We conclude that in the limit where one scales only the dilaton field, the effect of the one-loop string correction
on the NS5-instanton contributions is very non-trivial and captured by the parabolic cylinder functions,
but consistently with physical expectations it affects only the prefactor corresponding to the
determinant of fluctuations around the instanton.

\providecommand{\href}[2]{#2}\begingroup\raggedright\endgroup

%\bibliography{combined}
%\bibliographystyle{utphys}

\end{document}